\documentclass[reprint, prd, amsmath, amssymb, superscriptaddress, nofootinbib]{revtex4-2}
\usepackage{graphicx,color,float,tabularx,siunitx}
\usepackage[shortlabels]{enumitem}%
\usepackage[caption=false]{subfig}
\usepackage{placeins}
\usepackage{rviewport}
\definecolor{colorLink} {rgb}{0,0,0.8} 
\definecolor{colorCite} {rgb}{0,0,0.8} 
\definecolor{colorURL} {rgb}{0,0,0.8} 
\usepackage[colorlinks=true,linktocpage=true,linkcolor=colorLink,citecolor=colorCite,urlcolor=colorURL]{hyperref}

\usepackage[sort&compress]{natbib}

\begin{document}

\title{Probing the environments surrounding ultrahigh energy cosmic ray accelerators\\and their implications for astrophysical neutrinos}

\author{Marco Stein Muzio}
\email{msm659@nyu.edu}
\affiliation{Center for Cosmology and Particle Physics, Department of Physics, New York University, 726 Broadway, New York, New York 10003, USA}
\author{Glennys R. Farrar}
\email{gf25@nyu.edu}
\affiliation{Center for Cosmology and Particle Physics, Department of Physics, New York University, 726 Broadway, New York, New York 10003, USA}
\author{Michael Unger}
\email{Michael.Unger@kit.edu}
\affiliation{Institut f\"ur Astroteilchenphysik, Karlsruher Institut f\"ur Technologie, Karlsruhe 76344, Germany}

\date{\today}

\begin{abstract}
We explore inferences on ultrahigh energy cosmic ray (UHECR) source environments -- constrained by the spectrum and composition of UHECRs and non-observation of extremely high energy neutrinos -- and their implications for the observed high energy astrophysical neutrino spectrum. We find acceleration mechanisms producing power-law CR spectra~$\propto E^{-2}$ are compatible with UHECR data, if CRs at high rigidities are in the quasi-ballistic diffusion regime as they escape their source environment. Both gas-dominated and photon-dominated source environments are able to account for UHECR observations, however photon-dominated sources give a better fit. Additionally, gas-dominated sources are in tension with current neutrino constraints. Accurate measurement of the neutrino flux at $\sim 10$ PeV will provide crucial information on the viability of gas-dominated sources, as well as whether diffusive shock acceleration is consistent with UHECR observations. We also show that UHECR sources are able to give a good fit to the high energy portion of the astrophysical neutrino spectrum, above $\sim$ PeV. This common origin of UHECRs and high energy astrophysical neutrinos is natural if air shower data is interpreted with the \textsc{Sibyll2.3c} hadronic interaction model, which gives the best-fit to UHECRs and astrophysical neutrinos in the same part of parameter space, but not for EPOS-LHC. 
\end{abstract}	

\maketitle

\section{Introduction} \label{sec:intro}

\par
Discovering the origin of ultrahigh energy cosmic rays (UHECRs), $E \gtrsim 10^{18}$ eV ($1$ EeV), has been a longstanding challenge due to their incredibly low flux, their uncertain composition, and their deflection in both Galactic and extragalactic magnetic fields \cite{Anchordoqui18}. However high energy and UHE neutrinos, which do not suffer magnetic deflections or significant energy losses, are natural signals of interactions of UHECRs (and their secondary CRs) with photon fields and gas during their propagation from the accelerator to Earth. The potential of neutrinos to provide multimessenger observations to help discover UHECR origins has been underscored by the recent coincident observations of a high energy neutrino (IC-170922A) with the flaring blazar TXS 0506+056 \cite{TXSObservation} and another with the tidal disruption event TDE AT2019dsg \cite{Stein+20}.

\par
In this paper, we build on the approach of Unger, Farrar, and Anchordoqui (UFA 2015, \cite{UFA15}) and its elaboration in \cite{MUF19}, to investigate constraints on UHECR sources which can be inferred from the latest multimessenger data, using a realistic but flexible UHECR source model with minimal reliance on any specific astrophysical scenario. UFA15 proposed such a model for UHECR sources, taking into account photohadronic interactions of UHECRs propagating through the source environment after acceleration (potentially delayed in their escape due to diffusion in a turbulent magnetic field) and before extragalactic propagation. This model provides a natural explanation for the observed UHECR spectrum and its composition, and enabled multimessenger constraints on UHECR sources to be investigated in \cite{MUF19}. We consider here the effects of gas in the environment surrounding the source and the ability of UHECR sources to explain the highest energy astrophysical neutrinos.\footnote{Earlier studies have also considered that UHECRs and
astrophysical neutrinos may have a common origin. Qualitative fits to the astrophysical neutrino spectrum consistent with UHECR observations in the context of photodisintegration models are given in \cite{Giacinti+15,Kachelriess+17,Yoshida+20} while for specific astrophysical scenarios see \cite{Biehl+17b,Fang+17,Boncioli+18,Guepin+17,Zhang+18,Rodrigues+20}. Other studies have explored neutrino backgrounds at higher energies consistent with UHECR data \cite{Biehl+17a,Globus+17,AlvesBatista+18,Heinze+20} and have set upper-bounds on the flux of neutrinos produced by UHECRs \cite{Waxman+98}.} Like \cite{UFA15,MUF19}, our description is agnostic to the acceleration process or specific astrophysical source type. 

\begin{figure*}[htbp!]
	\centering
    \includegraphics[width=\textwidth]{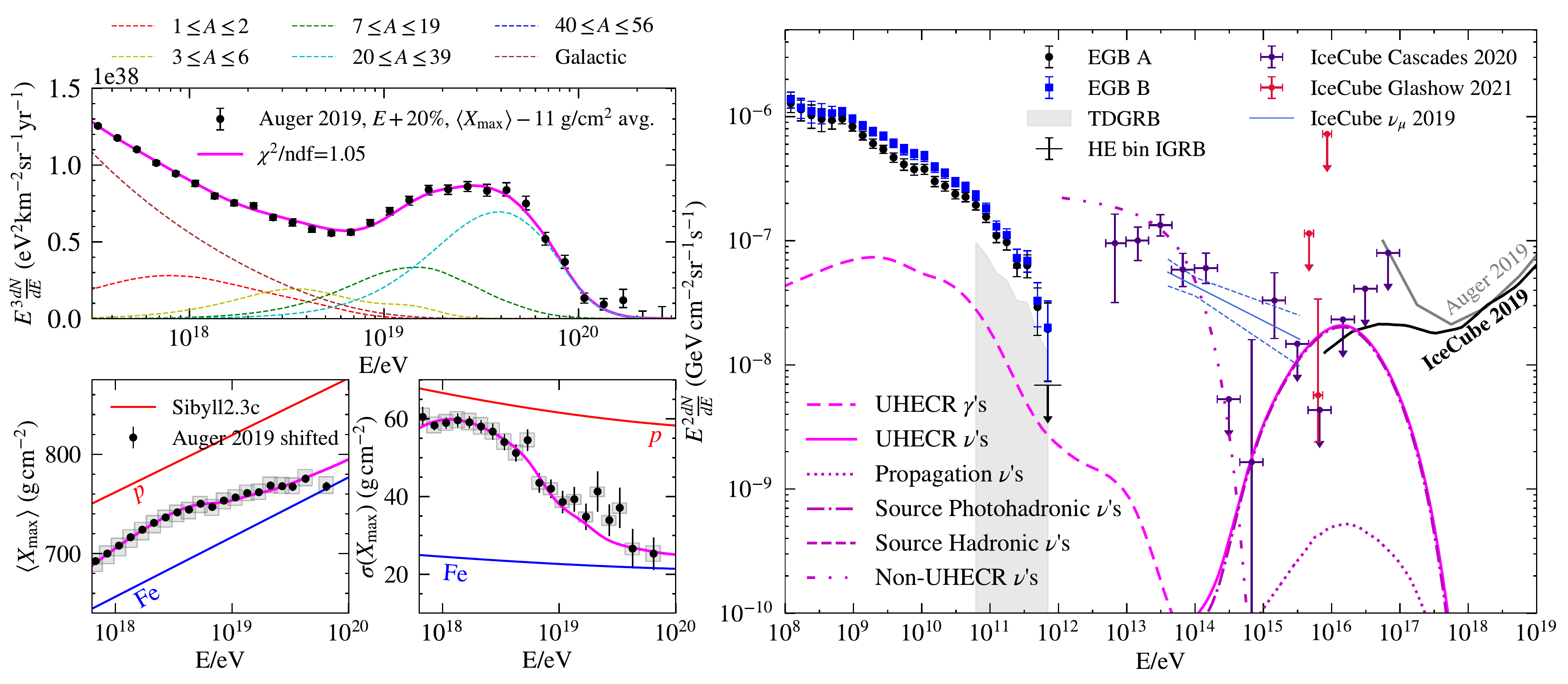}
	\caption{Predictions of the UHECR source model producing the best description of the astrophysical neutrino flux for \textsc{Sibyll2.3c}. \textbf{Left:} The CR predictions for spectrum (top) and composition (bottom) compared to shifted Auger observations \cite{Aab+20a,Aab+20b,Verzi20,Yushkov20}, as detailed in Section \ref{sec:data}. The red and blue solid lines show the $\langle X_\mathrm{max}\rangle$ and $\sigma (X_\mathrm{max})$ predictions of \textsc{Sibyll2.3c} for pure proton and iron models. \textbf{Right:} The neutrino and gamma-ray predictions for this model (solid and dashed lines, respectively). The total neutrino flux due to UHECRs (``UHECR $\nu$'s'', solid magenta) is broken down by origin: UHECR interactions during extragalactic propagation (``Propagation $\nu$'s'', dotted dark magenta), UHECR photohadronic and hadronic interactions in the source (``Source $\nu$'s'', dot-dashed and dashed dark magenta, respectively). Neutrinos originating from a source other than UHECRs (``Non-UHECR $\nu$'s'', dot-dot-dashed dark magenta) are also shown. The observed and inferred values of the extragalactic gamma-ray flux \cite{Ackermann+14}, astrophysical neutrino fluxes \cite{IceCubeCascades20,IceCubeMuon19}, flux measurements from the Glashow event \cite{IceCubeGlashow2021}, and upper-bounds on the EHE cosmic neutrino flux from IceCube \cite{Aartsen+18,IceCubeGlashow2021} (black) and Auger \cite{Aab+2019} (grey) are shown. Data points are as detailed in the text.}
	\label{fig:astroNuBestFits_sibyll}
\end{figure*}

\par
We focus here on the following questions:
\begin{enumerate}[(a)]
\item How is the original ``UFA15" picture impacted by the presence of gas in the surroundings of the accelerator, in addition to -- or instead of -- a photon field?
\item Is the diffusive shock acceleration prediction for the power-law index of the spectrum of UHECRs emerging from their accelerator consistent with multimessenger data?
\item Can the highest energy neutrinos so far detected be produced by interactions of UHECRs with photons and/or gas in their source environment, given the strongly constrained UHECR spectrum and composition?
\end{enumerate}

\par
This paper is organized as follows. In Section \ref{sec:modeling} we review the model of UFA15 and detail the elaborations made for this study. Section \ref{sec:data} gives an overview of the relevant multimessenger data and the constraints we have adopted. Finally, in Section \ref{sec:results} we report our findings, including the degree to which gas- and photon-dominated source environments can be distinguished and the ability of UHECR sources to explain the high energy astrophysical neutrino flux. For the reader's orientation, Fig. \ref{fig:astroNuBestFits_sibyll} shows the UHECR model giving the best-fit to the astrophysical neutrino flux, among those giving a good fit to UHECR spectrum and composition data using the \textsc{Sibyll2.3c} hadronic interaction model, as discussed below.

\section{Model} \label{sec:modeling}

\subsection{Overview}

\par
To perform this analysis we extend the UFA15 framework by adding interactions with gas in the source environment (as described in Section \ref{sec:gasInts}), as well as making a variety of technical improvements to the analysis in \cite{UFA15}. Based on the results of \cite{MUF19}, we adopt a source evolution following the star-formation rate (SFR, \cite{Robertson+15}) and take a single-mass injection of CRs into the source environment.\footnote{Source evolutions which are stronger at high redshift generally result in a larger neutrino flux at Earth, but the quality of the fit to UHECR observations degrades and requires an extremely hard CR spectrum escaping the source \cite{MUF19,Aab+16}. Source evolutions with lower source densities at high redshift produce smaller neutrino fluxes at Earth and also give a worse fit to UHECR data when considering photodisintegration models, so are not explored in this study. However, UHECR data alone does not strongly constrain the source evolution. A more complex injection composition was found not to be needed \cite{MUF19}, so for simplicity we adopt single-mass injection.} We approximate the gas to be pure hydrogen since other components make up less than $10\%$ by number. We also introduce an alternate treatment of the energy dependence of the escape time, based on the behavior of diffusion coefficients and reflecting the finite size of sources. Details of our treatment of systematic uncertainties are given in Section \ref{sec:data}.

\par
A CR nucleus of energy $E$, mass $A$, and charge $Z$ has interactions with photons and gas at a rate $\tau^{-1}_\gamma(E,A)$ and $\tau^{-1}_g(E,A)$ when propagating in the source environment. These rates are specified by their cross sections, the photon spectral density distribution, and the gas density. Thus we can fully characterize UHECR interactions with photons and gas in a given source environment by knowing the parameters specifying the photon spectrum, and $\tau_\gamma$ and $\tau_g$ for a reference nucleus and energy. Following \cite{UFA15}, we adopt 10 EeV $^{56}$Fe as this reference.

\par
Only the total number of interactions prior to escape matters in the processing of nuclei injected by the accelerator, so only the ratio of the interaction and escape times is relevant for fixing the composition and energy spectra of CRs emerging from the source environment, given the injected composition and spectrum.\footnote{The overall normalization of the energy spectrum is set by the product of the UHECR luminosity per source and the number density of UHECR sources. However, this is unrelated to the interaction and escape times within the source environment and so can be fit independently to obtain the best-fit.} We denote these ratios for the reference nucleus by the model parameters $r_\mathrm{esc} \equiv \tau_\mathrm{esc}^\mathrm{ref}/\tau_\mathrm{int}^\mathrm{ref} = \langle N_\mathrm{int}^\mathrm{ref} \rangle$, the ratio of the escape and total interaction times, and $r_{g\gamma} \equiv \tau_g^\mathrm{ref} / \tau_\gamma^\mathrm{ref} = \langle N_\gamma^\mathrm{ref} \rangle / \langle N_g^\mathrm{ref} \rangle$, the ratio of the hadronic and photohadronic interaction times. 

\par
In previous work \cite{UFA15,MUF19}, the escape time was taken to be a power law in rigidity, $R\equiv E/Z$, so that $\tau_{\rm esc} = \tau^\mathrm{ref}_{\rm esc} (R/R_\mathrm{ref})^{\delta_{\rm esc}}$ with $\delta_\mathrm{esc}$ a free parameter of the model limited to be within $[-1,-1/3]$, covering the expected range from Kolmogorov to Bohm diffusion. Here we adopt a more detailed parameterization (described in Section \ref{sec:diffusion}) based on simulations of CR propagation in turbulent magnetic fields, e.g. \cite{Globus+07}: $\tau_\mathrm{esc}(R) = \frac{L^2}{6D(R)} + \frac{L}{c}$. This implies faster escape at high rigidities by capturing the quasi-ballistic regime. 

\par 
The interplay between the energy dependence of the interaction time, $\tau_\mathrm{int}(E)$, and the rigidity dependence of the escape time from the environment, $\tau_{\rm esc}(R)$, governs the energy dependence of the number of interactions before escape, as was noted in equation (4) of \cite{UFA15}. There, the focus was on how the high-pass filter mechanism creates the ankle in the UHECR spectrum and produces the observed light extragalactic UHECRs below the ankle. As noted in \cite{UFA15}, this mechanism leads to a hardening of the escaping CR spectrum. Thus the Auger combined fit to the spectrum and composition \cite{Aab+16} favoring an $\sim E^{-1}$ or harder spectrum escaping the source, does not necessarily mean that the accelerator produces such a hard spectrum. In principle, the fundamental spectrum from the accelerator could be much softer, e.g., $\sim E^{-2}$ as predicted by diffusive shock acceleration, with the low-energy component of any given $A$ being more depleted through interactions because of their longer residence time in the environment -- thus hardening the escaping spectrum for each composition.\footnote{An alternative mechanism to harden the accelerated spectrum is through magnetic suppression of the spectrum at low-rigidities due to the horizon induced by the extragalactic magnetic field \cite{Mollerach+13,Wittkowski17}.} 

\par
In this work we find that UHECR data is compatible with an accelerator spectrum $E^{-2}$, if the peak of that spectrum falls in the quasi-ballistic regime where the rapid change of escape time with energy leads to a strong hardening of the escaping flux. This requires source conditions such that high rigidity CRs are able to enter the quasi-ballistic diffusion regime. In this case, diffusive shock acceleration remains a possible mechanism for UHECR acceleration. As we shall show below, the viability of this scenario can be tested with accurate measurement of the neutrino flux at $\sim 10$ PeV. 

\par
A description of how our model's parameters can be related to astrophysical quantities will be given in a forthcoming paper.

\subsection{Incorporation of hadronic interactions in the source environment} \label{sec:gasInts}

\par
Photohadronic interactions in the source environment are accounted for as described in Appendix B and Appendix C of \cite{UFA15}, with the adjustment of the definition of the total interaction time to include hadronic interactions, so that

\begin{align} \label{eq:totIntDef}
  \tau^{-1}_\mathrm{int}(E,A) = \tau^{-1}_g(E,A) + \tau^{-1}_\gamma(E,A)
\end{align}

\noindent
is the total interaction time for a CR of mass $A$ and energy $E$, where $\tau_g$ is its hadronic interaction time and $\tau_\gamma$ is its photohadronic interaction time. 

\par
The photohadronic interaction time is the total interaction time due to both photopion (PP) and photodisintegration (PD) interactions with the ambient photon field,

\begin{align}
  \tau_\gamma^{-1}(E,A) = \tau_\mathrm{PP}^{-1}(E, A) + \tau_\mathrm{PD}^{-1}(E,A).
\end{align}

\begin{figure*}[htbp!]
    \begin{minipage}{0.49\linewidth}
	  \centering
	  \subfloat[\label{fig:timescales_sibyll}]{\includegraphics[width=\linewidth]{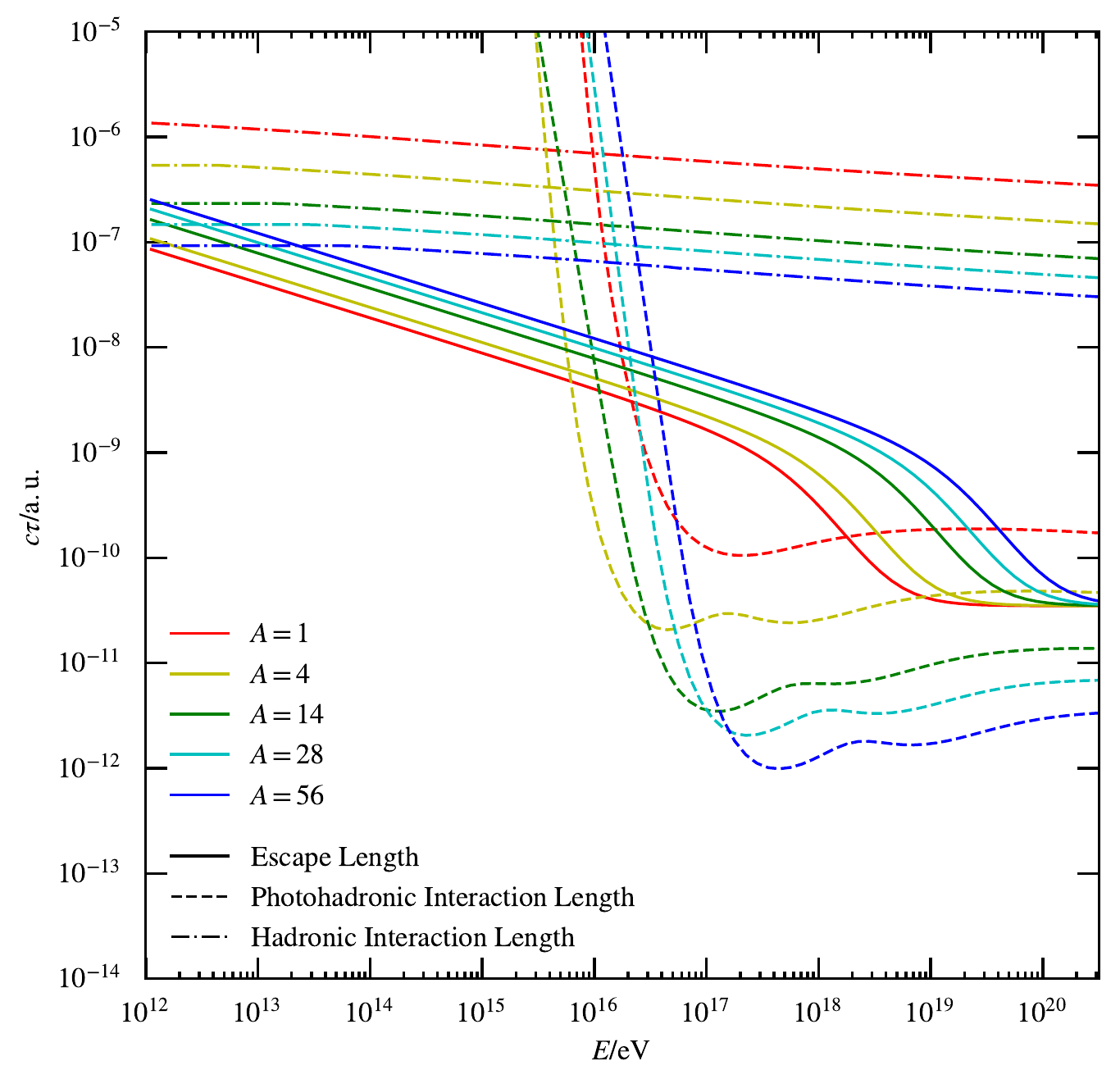}}
    \end{minipage}
    \begin{minipage}{0.49\linewidth}
	  \centering
	  \subfloat[\label{fig:timescales_epos}]{\includegraphics[width=\linewidth]{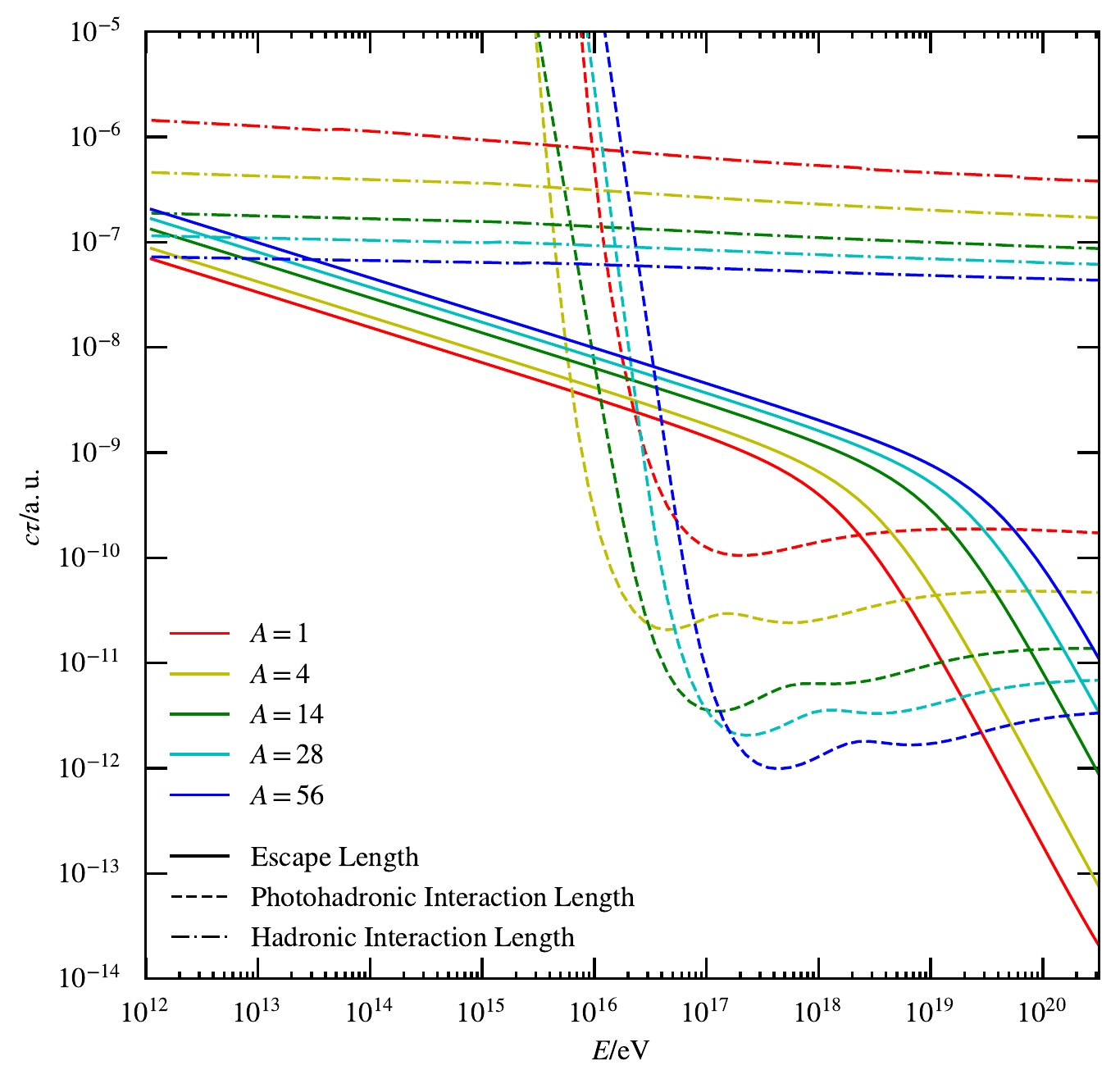}}
    \end{minipage}
    \caption{The escape (solid lines), photohadronic interaction (dashed lines), and hadronic interaction (dashed-dotted lines) lengths for different nuclear masses (colors) in the UHECR source model best reproducing the astrophysical neutrino flux for \textsc{Sibyll2.3c} (left) and \textsc{EPOS-LHC} (right). Hadronic interaction times were calculated under the respective hadronic interaction models. In particular while the interaction times are similar, the escape times in the two panels differ significantly at high energies, reflecting significantly different best-fit values of $r_\mathrm{size}= L/\lambda_c$. The corresponding fits are pictured in Figs. \ref{fig:astroNuBestFits_sibyll} and \ref{fig:astroNuBestFits_epos}, respectively.}
    \label{fig:timescales}
\end{figure*}

\noindent
The relation between the interaction times for each of these interaction channels and the ambient photon field are given below. 

\par
Throughout the present work we consider either a pure or generalized black-body photon spectral density distribution\footnote{
In \cite{MUF19}, we considered both black-body and broken power-law parametrizations of the photon spectral density distribution in the source environment and found that both gave comparable fits to the UHECR data. Black-body spectra, though, produced fewer extremely high energy cosmic neutrinos, allowing more conservative constraints to be set. The qualitative conclusions are expected to be similar for black-body and broken power-law parametrizations \cite{UFA15,Fiorillo+21}.} of the form

\begin{align} \label{eq:BB_def}
  n(\varepsilon) = n_0 f_\mathrm{BB}(\varepsilon)
\end{align}

\noindent
where $n_0$ is the photon density in units of $I_\mathrm{BB} \equiv \frac{16\pi\zeta(3)}{(hc)^3} \left(k T\right)^3 $, the black-body photon number density, and $f_\mathrm{BB}$ is the spectral density distribution of a black-body (BB) photon field with temperature $T$ given by

\begin{align}
  f_\mathrm{BB}(\varepsilon) &= \frac{8\pi}{(hc)^3} \frac{\varepsilon^2}{\mathrm{e}^\frac{\varepsilon}{kT}-1}.
\end{align}

\noindent
The generalized black-body spectral density \eqref{eq:BB_def} corresponds to the spectral density of a grey-body for $n_0 < 1$. With this definition, the physical number density of the photon spectral density distribution $n(\varepsilon)$ is given by $n_\gamma = n_0 I_\mathrm{BB}$.

\par
The interaction time for each of the photohadronic interaction channels, $i=$ PD or PP, is given by

\begin{align}
  \tau_i^{-1}(E, A) &= \frac{c}{2} \int_0^\infty \mathrm{d}\varepsilon\frac{n(\varepsilon)}{\gamma^2\varepsilon^2} \int_0^{2\gamma\varepsilon} \mathrm{d}\varepsilon' \varepsilon' \sigma_{A\gamma, i}(\varepsilon') \\
                    &= n_0 \frac{c}{2} \int_0^\infty \mathrm{d}\varepsilon\frac{f_\mathrm{BB}(\varepsilon)}{\gamma^2\varepsilon^2} \int_0^{2\gamma\varepsilon} \mathrm{d}\varepsilon' \varepsilon' \sigma_{A\gamma, i}(\varepsilon') \\
					&= n_0 \tau_{\mathrm{BB},i}^{-1}(E, A, T) \label{eq:gammatab}
\end{align}

\noindent
where $\sigma_{A\gamma,i}$ is the photonuclear cross section for photohadronic interaction channel $i$ (provided by \textsc{CRPropa} \cite{AlvesBatista+16} based on TALYS and SOPHIA to appropriately handle the different photon energy regimes), $\gamma$ is the Lorentz factor, and $\tau_{\mathrm{BB},i}$ is the interaction time for a black-body distribution with temperature $T$.

\par
In order to account for hadronic interactions we must assume a hadronic interaction model (HIM), as hadronic interactions within the source environment occur at similar center-of-mass energies as in extensive air showers initiated by UHECRs on Earth. Once we have assumed a particular HIM, labeled $m$, the hadronic interaction time is determined by the density of gas $n_g$ in the source environment and the $pA$ cross section $\sigma_g^m (E,A)$, for a collision between a proton at rest and CR of energy $E$ in that model:

\begin{align}
  \tau_g^{-1}(E,A) = n_g \sigma^m_g(E, A) c.
\end{align}

\par
Particle production in these interactions also depends upon the assumed HIM. Using \textsc{CRMC} v1.6.0 \cite{CRMC}, we built interaction matrices for the \textsc{EPOS-LHC} \cite{Pierog+13} and \textsc{Sibyll2.3c} \cite{Fedynitch+18} HIMs.\footnote{Throughout this work, the HIM assumed for calculation of interactions in the source environment always matched the HIM used to interpret air shower observations.} Simulating the collision between a CR of nucleus $A$ and energy $E_j$ with a proton at rest, we generated an interaction matrix $I_{ij}^{sA,m}$ by tabulating all the secondaries of type $s$ (e.g. $\nu$, $\pi$, $\gamma$, $A'$) and energy $E_i$ predicted by HIM $m$. Pions and neutrons were treated as stable for this purpose,\footnote{Here we assume the energy loss time due to synchrotron radiation exceeds either the decay time of the $\pi\rightarrow\mu\rightarrow\nu$ chain or the escape time of pions and muons. See e.g. \cite{Baerwald+11} for a discussion of cooling-damped sources. In particular, our results here are robust to cooling as will be shown in a forthcoming paper.} in order to treat these particles in a consistent way with those produced in photohadronic interactions (as described in \cite{UFA15}), and all other particles were decayed before being tabulated. The final interaction matrices give the average number of secondaries of energy $E_i$ produced in a collision. With the interaction matrices in hand, one can easily compute the total number of secondaries $s$ produced in logarithmic interval $[\lg{E_i}, \lg{E_i}+\mathrm{d}\lg{E_i})$ by a flux of CRs as

\begin{align}
  \frac{\mathrm{d}N^{s}}{\mathrm{d}\lg{E_i}} = \sum_{\substack{j>i,A}} I^{sA,m}_{ij} f_g(A, E_j) f_\mathrm{int}(A, E_j) \frac{\mathrm{d}N^{A}}{\mathrm{d}\lg{E_j}} \left|\frac{\mathrm{d}\lg{E_j}}{\mathrm{d}\lg{E_i}} \right|
\end{align}

\noindent
where $f_g \equiv \left(1+ \tau_g/\tau_\gamma \right)^{-1}$ is the fraction of interactions which are hadronic, and $f_\mathrm{int} \equiv \left(1+ \tau_\mathrm{int}/\tau_\mathrm{esc} \right)^{-1}$ is the fraction of CRs which interact. In the above formula, we have neglected reinteraction of secondary CRs for simplicity and clarity, but in our full calculation reinteractions are accounted for. 

\par
It is important to note that hadronic interactions always dominate at low enough energies such that the $\gamma$-CR center-of mass-energy is too low to photopion produce or excite the giant dipole resonance. This is reflected in the energy dependence of the interaction times, as Fig. \ref{fig:timescales} shows. Thus even in photon-dominated environments (meaning environments where UHECRs primarily interact photohadronically), astrophysical neutrinos may still be the product of hadronic interactions with gas, given a sufficient gas density and number of interactions. On the other hand, CRs in gas-dominated sources primarily interact hadronically at both high and low energies.

\subsection{CR diffusion and escape from the source environment} \label{sec:diffusion}

In previous analyses \cite{UFA15} and \cite{MUF19}, the rigidity dependence of the diffusion coefficient was modeled as a single power law in rigidity: $D(R) \propto R^{-\delta_\mathrm{esc}}$ and $\tau_\mathrm{esc} \propto R^{\delta_\mathrm{esc}}$. This is a good approximation for CRs that are in the diffusive regime for all the relevant rigidities. For a more realistic treatment we use a fit to the diffusion simulation of \cite{Globus+07} with the functional form suggested in \cite{Harari+13} (leading to very similar numerical results as found therein):

\begin{align} \label{eq:GAP+BohmR}
	D(R) &= \frac{c\lambda_c}{6\pi} \left[ \left(\frac{R}{R_\mathrm{diff}}\right)^{1/3} +              \frac{1}{2} \left(\frac{R}{R_\mathrm{diff}}\right) + \frac{2}{3}                           \left(\frac{R}{R_\mathrm{diff}}\right)^2 \right],
\end{align}

\noindent
where $r_L(R) \equiv 1.1\text{ kpc} \left(\frac{\text{$\mu$G}}{B}\right) \left(\frac{R}{\text{EV}}\right)$ is the Larmor radius, $\lambda_c$ is the coherence length of the magnetic field, $B$ is its RMS field strength, and $R_\mathrm{diff}$ is the characteristic rigidity scale of diffusion, defined by $2\pi r_L(R_\mathrm{diff}) \equiv \lambda_c$. For convenience, we define the dimensionless diffusion coefficient, $d(R) \equiv \left(R/R_\mathrm{diff}\right)^{1/3} + \frac{1}{2} \left(R/R_\mathrm{diff}\right) + \frac{2}{3} \left(R/R_\mathrm{diff}\right)^2$, so that $D(R) = \frac{c\lambda_c}{6\pi}d(R)$.

\par
CRs of rigidity $R$ that enter the diffusion regime have an escape time given by $\tau_\mathrm{esc} = \frac{L^2}{6D(R)}$. However, in reality the source may not be large enough for all CRs to enter the diffusion regime before escaping the source, especially those at high rigidities with large Larmor radii. The source's physical size $L$ then imprints a minimum escape time of $L/c$. To account for this effect we adopt the following parametrization for the escape time, which has been verified against simulation data shown in \cite{Globus+07}:

\begin{align} \label{eq:escTime_def}
    \tau_\mathrm{esc}(R) = \frac{L^2}{6D(R)} + \frac{L}{c}~.
\end{align}

\noindent
This treatment imposes that the escape time be at least as large as the crossing time.

\par
The escape time for a nucleus of rigidity $R$ can then be calculated in terms of the escape time of the reference nucleus as 

\begin{align}
  \tau_\mathrm{esc}(R) &= \tau_\mathrm{esc}^\mathrm{ref} \frac{\frac{L^2}{6D(R)} + \frac{L}{c}}{\frac{L^2}{6D(R_\mathrm{ref})} + \frac{L}{c}} \nonumber \\
                       &= \tau_\mathrm{esc}^\mathrm{ref} \frac{\frac{\pi}{d(R)}\frac{L}{\lambda_c} +1}{\frac{\pi}{d(R_\mathrm{ref})}\frac{L}{\lambda_c} +1} \nonumber \\
                       &= \tau_\mathrm{esc}^\mathrm{ref} \frac{\frac{\pi r_\mathrm{size}}{d(R)} +1}{\frac{\pi r_\mathrm{size}}{d(R_\mathrm{ref})} +1},
\end{align}

\noindent
where $R_\mathrm{ref} = 10/26\text{ EV} \simeq 0.38\text{ EV}$ is the rigidity of the reference nucleus, $\tau_\mathrm{esc}^\mathrm{ref}$ is its escape time, and we have introduced the model parameter $r_\mathrm{size} \equiv L/\lambda_c$. 

\par
We find that in order for a soft $E^{-2}$ acceleration spectrum to be adequately hardened to account for UHECR data, high rigidity CRs must be in the $\tau_\mathrm{esc} \sim E^{-2}$ regime. This implies that $r_\mathrm{size} \gg 1$ in order for conventional acceleration mechanisms to explain UHECR observations.

\begin{figure*}[htpb!]
	\centering
	\begin{minipage}{0.497\linewidth}
	  \centering
	  \subfloat[\label{fig:Nsigma_BB_sibyll}]{\includegraphics[width=\linewidth]{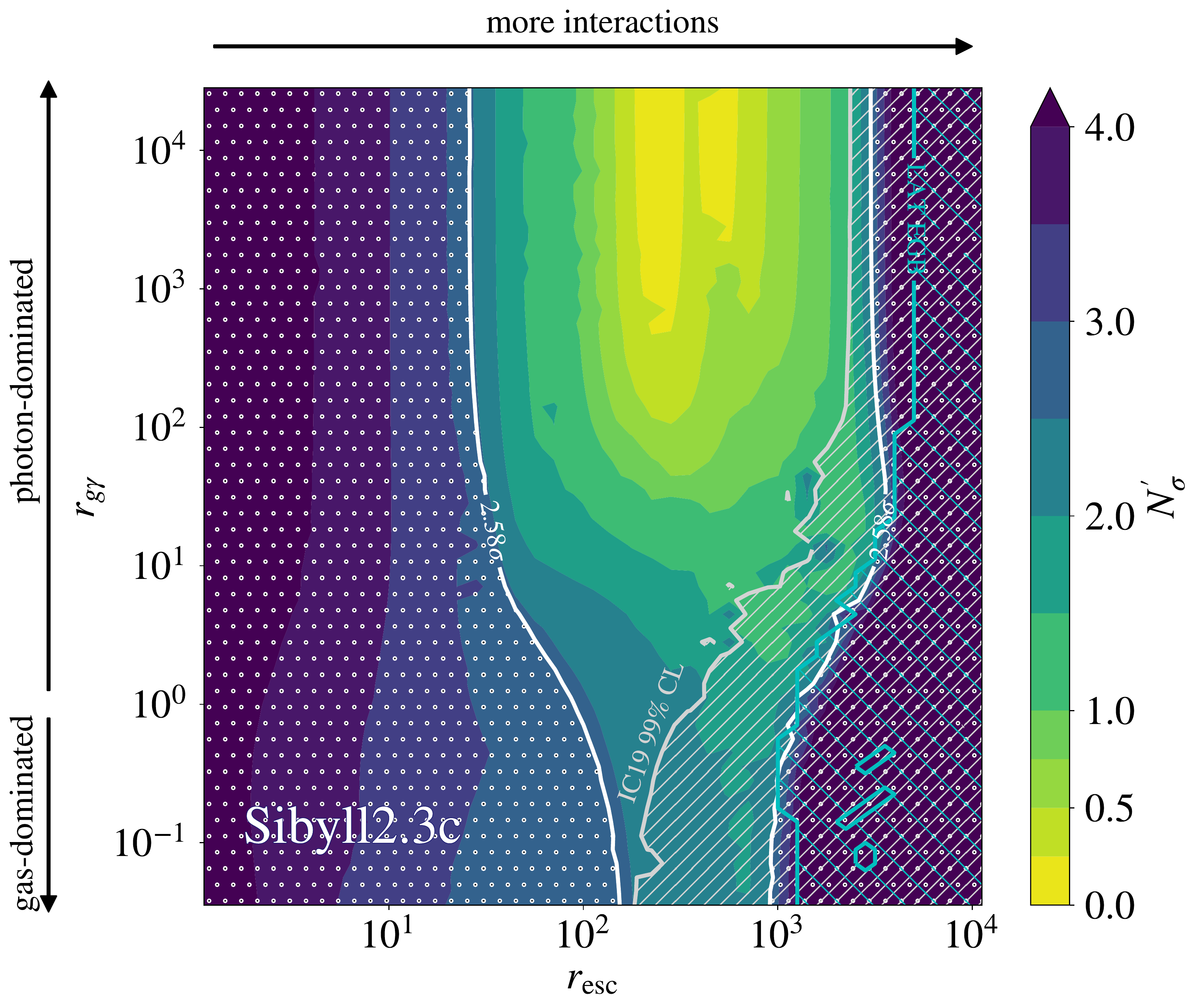}}
    \end{minipage}
    \begin{minipage}{0.497\linewidth}
	  \centering
	  \subfloat[\label{fig:Nsigma_BB_epos}]{\includegraphics[width=\linewidth]{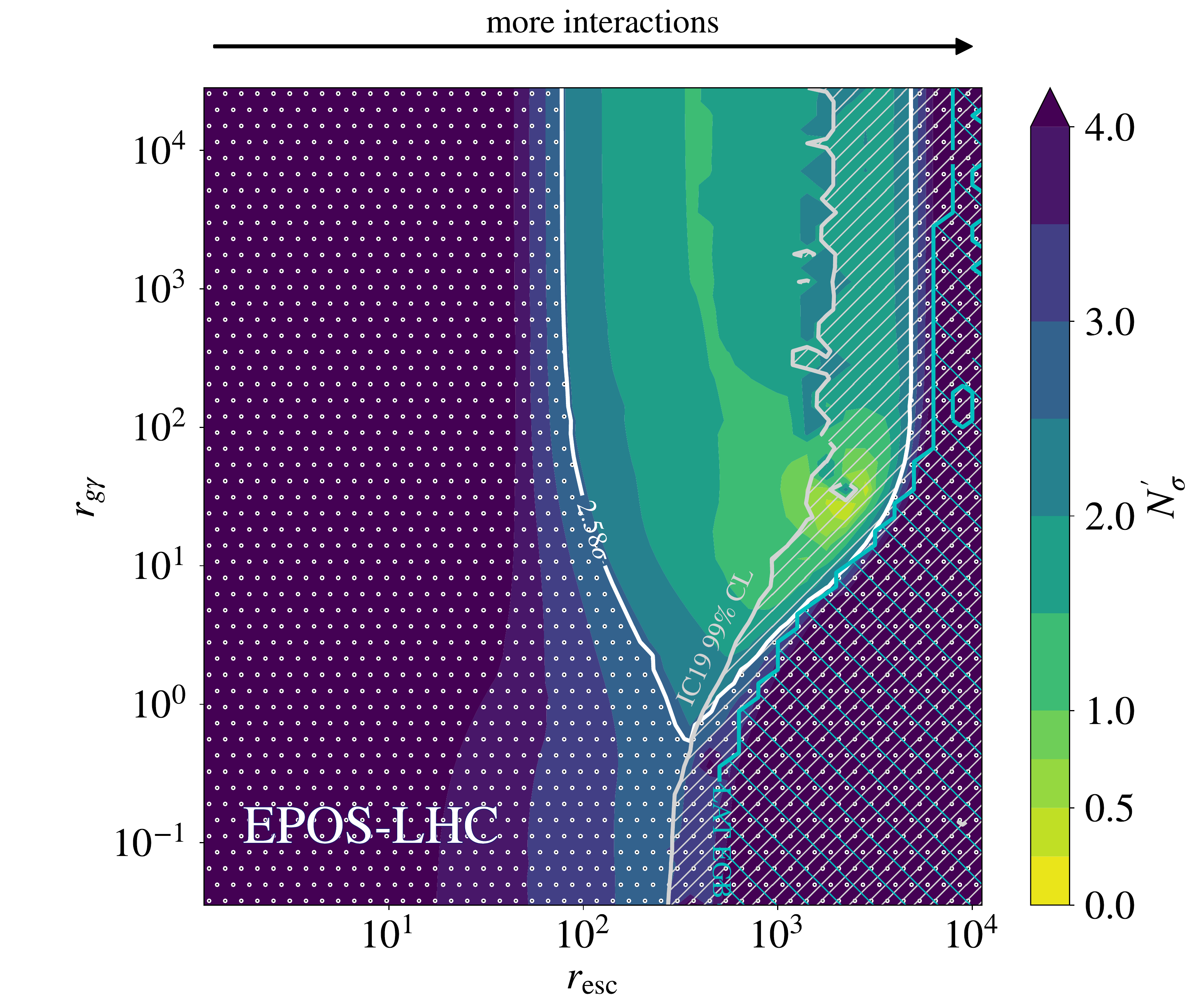}}
    \end{minipage}
    \caption{$N_\sigma'$, the number of standard deviations improvement of the global UHECR best-fit relative to the local UHECR best-fit, as a function of model parameters $r_\mathrm{esc}\equiv \tau_\mathrm{esc}^\mathrm{ref}/\tau_\mathrm{int}^\mathrm{ref}$ and $r_{g\gamma} \equiv \tau_g^\mathrm{ref} / \tau_\gamma^\mathrm{ref}$ for \textsc{Sibyll2.3c} (left) and \textsc{EPOS-LHC} (right). Contours mark $99\%$ confidence level (CL) exclusion regions based on UHECR data (white contour and dotted region), non-observation of neutrinos above $10^{15.9}$ eV (grey contour and hatched region), and upper-bounds on the extragalactic gamma-ray flux (cyan contour and hatched region). The global best-fit points ($N'_\sigma =0$) correspond to $\chi^2/ndf$'s of $58.5/58$ and $106.2/58$ for \textsc{Sibyll2.3c} and \textsc{EPOS-LHC}, respectively.}
    \label{fig:Nsigma_BB}
\end{figure*} 

\section{Data \& Methodology} \label{sec:data}

\par
Our model is fit to the UHECR spectrum and composition data of Auger \cite{Aab+20a,Aab+20b,Verzi20,Abreu+13,Aab+14a,Aab+14b,Yushkov20}, applying a $+0.8$-bin shift of the Auger energy scale ($+20\%$, slightly larger than the quoted systematic uncertainty of $14\%$ \cite{Verzi13}). We also shift $\langle X_\mathrm{max} \rangle$ by $-10$ g/cm$^2$ on average, following the energy dependence quoted in \cite{Aab+14a}. These shifts were determined via a preliminary study by scanning over the energy and $\langle X_\mathrm{max} \rangle$ shifts and fixing their values to the combination which enabled the best-fit to the data. To assess the goodness-of-fit we compute a combined $\chi^2$ to the spectrum data and $\langle X_\mathrm{max} \rangle$ and $\sigma\left(X_\mathrm{max}\right)$ data mapped into $\langle \ln{A} \rangle$ and $\mathrm{V}(\ln{A})$ using the parametrization of \cite{Abreu+13},

\begin{align}
    \chi^2 = &\displaystyle\sum_i^{N_\mathrm{spec}} \frac{(J_{m,i}-J_i)^2}{\sigma_{J,i}^2} + \displaystyle\sum_j^{N_\mathrm{comp}} \frac{(\langle \ln{A} \rangle_{m,j} - \langle \ln{A} \rangle_j )^2}{\sigma_{\langle \ln{A} \rangle,j}^2} \nonumber \\
    &+ \displaystyle\sum_j^{N_\mathrm{comp}} \frac{(\mathrm{V}(\ln{A})_{m,j} - \mathrm{V}(\ln{A})_j)^2}{\sigma_{\mathrm{V}(\ln{A}),j}^2},
\end{align}

\noindent
where $N_\mathrm{spec}$ and $N_\mathrm{comp}$ are the number of data points in the spectrum and composition, respectively, $x_{m,i}$ denotes the model prediction of quantity $x$ (flux $J$, mean logarithmic mass $\langle\ln{A}\rangle$ or its variance $\mathrm{V}$) at energy bin $i$, and errors include both systematic and statistical errors for the spectrum \cite{Aab+20a,Aab+20b,Verzi20} and statistical errors for the mean and variance on $\ln{A}$ \cite{Abreu+13,Aab+14a,Aab+14b,Yushkov20}. For spectral energy bins above the highest-energy data point, we follow \cite{Baker+83} by adding an additional $2n_i$ to the $\chi^2$, where $n_i$ is the expected number of observed events predicted by the model in energy bin $i$ given the exposure of the dataset. Our final figure of merit is $\chi^2_\mathrm{CR} = \chi^2 + 2\sum_i n_i$, where $i$ runs over energy bins above the highest-energy data point in the spectrum.

\begin{figure*}[htpb!]
	\centering
	\begin{minipage}{0.49\linewidth}
	  \centering
	  \subfloat[\label{fig:gammaInj_BB_sibyll}]{\includegraphics[width=\linewidth]{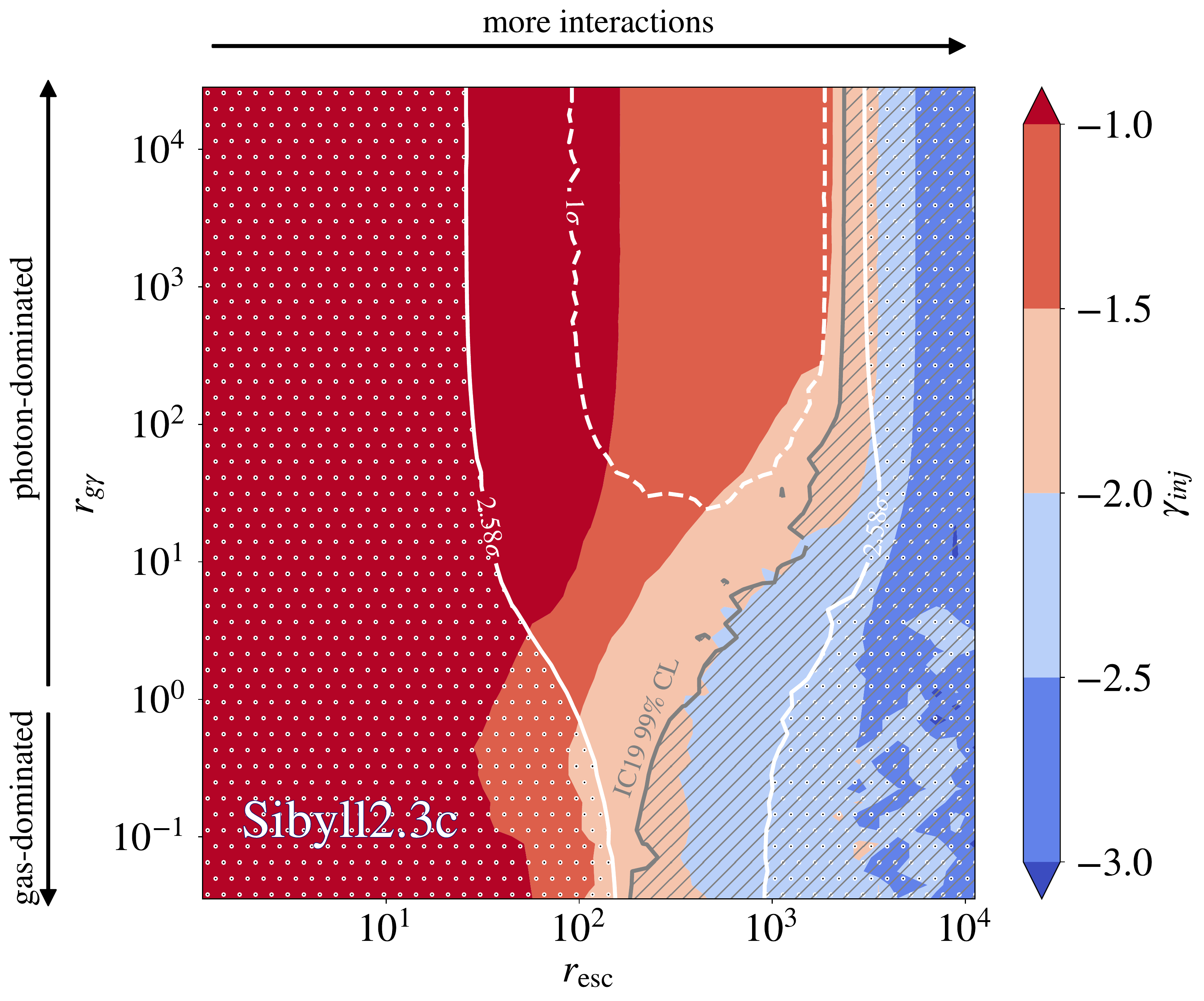}}
    \end{minipage}
    \begin{minipage}{0.49\linewidth}
	  \centering
	  \subfloat[\label{fig:gammaInj_BB_epos}]{\includegraphics[width=\linewidth]{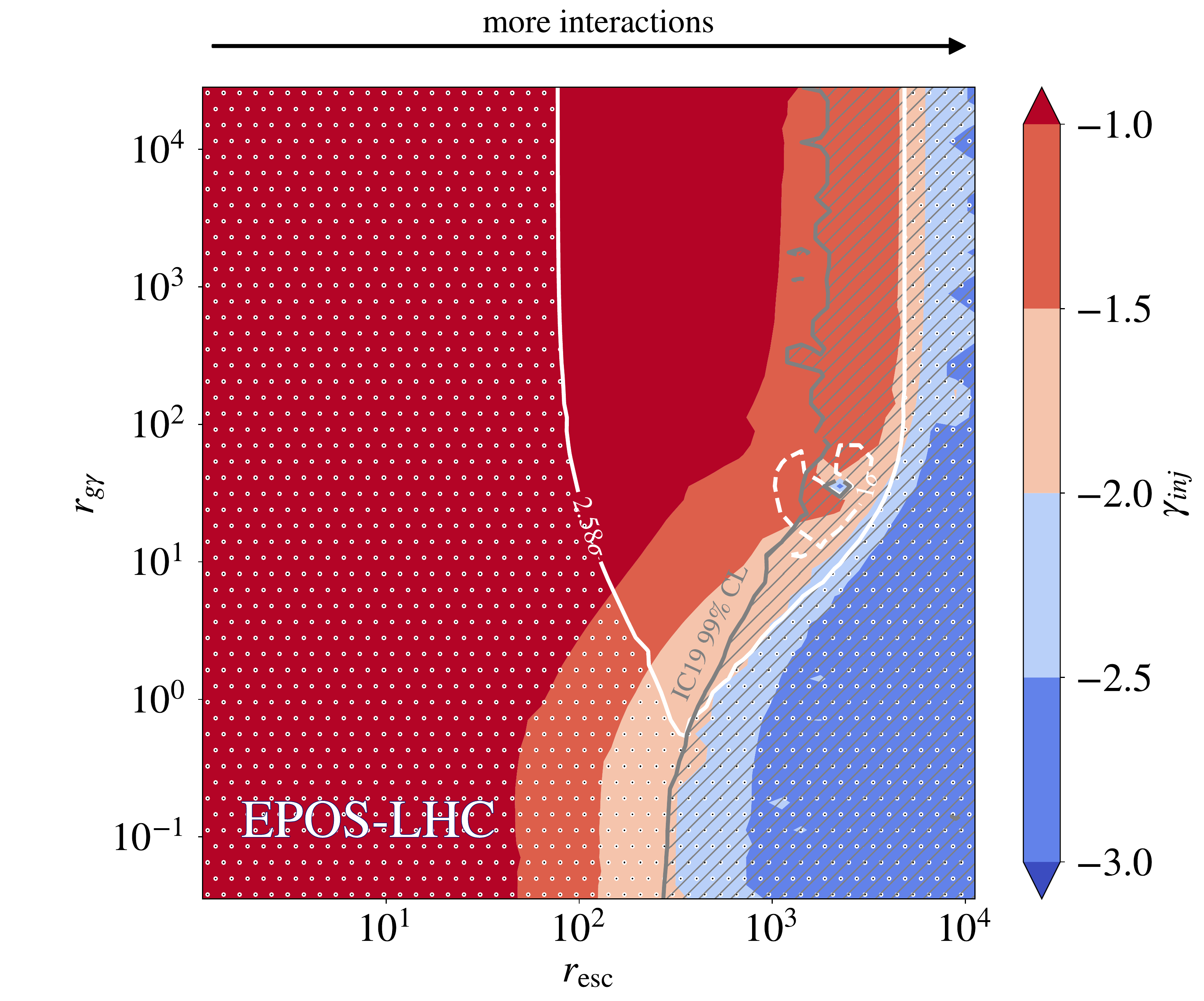}}
    \end{minipage}
    \caption{Spectral index $\gamma_\mathrm{inj}$ of CRs injected into the source environment, $E^{\gamma_\mathrm{inj}}$, as a function of model parameters $r_\mathrm{esc}\equiv \tau_\mathrm{esc}^\mathrm{ref}/\tau_\mathrm{int}^\mathrm{ref}$ and $r_{g\gamma} \equiv \tau_g^\mathrm{ref} / \tau_\gamma^\mathrm{ref}$ for \textsc{Sibyll2.3c} (left) and \textsc{EPOS-LHC} (right). Contours mark $99\%$ CL exclusion regions based on UHECR data (white contour and dotted region) and non-observation of neutrinos above $10^{15.9}$ eV (grey contour and hatched region).}
    \label{fig:gammaInj_BB}
\end{figure*} 

\par
We follow the PDG \cite{PDG, Rosenfeld75} defining the number of sigma from the best-fit as $N_\sigma' = S^{-1} \sqrt{\chi^2_\mathrm{model} - \chi^2_\mathrm{min}}$, where $S = \sqrt{\chi^2_\mathrm{min}/N_\mathrm{dof}}$ is the scale factor introduced in \cite{Rosenfeld75} to enlarge the uncertainties to account for a $\chi^2_\mathrm{min}/N_\mathrm{dof} > 1$, $\chi^2_\mathrm{model}$ is the $\chi^2$ for a given model, $\chi^2_\mathrm{min}$ is the $\chi^2$ of the best-fit model (under the assumption of that HIM), and $N_\mathrm{dof}$ is the number of degrees of freedom. We adopt the condition $N_\sigma'< 2.58$, corresponding to $99\%$ CL, when placing constraints based on UHECR data.

\par
We also place constraints using bounds on extremely high energy (EHE) neutrinos ($E_\nu > 10^{15.9}$ eV) from IceCube \cite{Aartsen+18}, updated to reflect the newly observed Glashow event \cite{IceCubeGlashow2021}. Models which predict $N_\nu^{\rm EHE} > 4.74$ are excluded at the $99\%$ CL, since IceCube has observed no neutrinos at these energies and this energy range is expected to be background free \cite{Feldman+97}. We allow fits that are within $2.58\sigma$ of the best-fit to CRs to stray from the local best-fit parameters, as long as they respect $N_\nu^{\rm EHE} < 4.74$, so as to determine if any choice of parameters is compatible with both sets of data. This process can introduce \textit{apparent} sampling artifacts into the results, e.g. along the grey contour in Fig. \ref{fig:Nsigma_BB_epos}.

\par
Finally, we consider the ability of current extragalactic gamma-ray data to constrain our model's parameters. For this we consider the extragalactic gamma-ray background (EGB) measured by the \textit{Fermi}-Large Area Telescope (LAT) \cite{Ackermann+14}. In order to place conservative bounds we adopt the published EGB spectrum under the assumption of LAT's Galactic foreground model B, the foreground model which attributes the least Galactic emission to the measured gamma-ray flux. We consider a model to be excluded if its predicted gamma-ray flux exceeds the EGB model B measurement by more than the error bar in any energy bin. In plots we also give an estimate of the truly diffuse gamma-ray background (TDGRB) -- as described in Section 2 of \cite{MUF19} -- and the flux measured in the highest energy bin of the isotropic gamma-ray background (IGRB) model B by LAT \cite{Ackermann+14}. As is shown in Fig. \ref{fig:Nsigma_BB_sibyll}, gamma-ray data is weakly constraining, only ruling out models also excluded by bounds on EHE neutrinos. Thus we set aside the gamma-ray constraints for the rest of our analysis.

\section{Results} \label{sec:results}

\subsection{Implications of gas in the source environment} \label{sec:ImpsOfGas}

\par
To a good approximation hadronic interactions partially disintegrating a nucleus preserve the energy-per-nucleon of the primary CR. This feature allows CR interactions with gas to realize the mechanism explored in \cite{UFA15}, and expanded upon in \cite{MUF19}, for explaining the observed UHECR spectrum and composition. Fits in Fig. \ref{fig:Nsigma_BB} show that both gas- and photon-dominated environments can give an adequate accounting of UHECR data. This figure shows how the quality of fit to UHECR data changes with $r_\mathrm{esc}$, controlling the average number of interactions CRs undergo before escape, and $r_{g\gamma}$, controlling whether the source environment is gas-dominated (small values) or photon-dominated (large values). Contours in this figure show the regions excluded by CR (white dotted region), EHE neutrino (grey hatched region), and gamma-ray (cyan hatched region) data. Figure \ref{fig:Nsigma_BB_sibyll} shows that while \textsc{Sibyll2.3c} is able to describe the UHECR data even for gas-dominated sources, it prefers photon-dominated sources. \textsc{EPOS-LHC}, though, prefers sources where photon interactions are more comparable to gas interactions, as can be seen in Fig. \ref{fig:Nsigma_BB_epos}. We also note that, in all cases, small values of $r_{esc}$ provide a poor fit to UHECR data since the injected UHECR spectrum must undergo a sufficient number of photodissociation interactions in order to account for the composition observed at Earth, in particular the position of the protonic component. Otherwise the composition of injected UHECRs must be fine tuned. 

\begin{figure*}
    \begin{minipage}{0.49\linewidth}
	  \centering
	  \subfloat[\label{fig:astroChi2_BB_sibyll}]{\includegraphics[width=\linewidth]{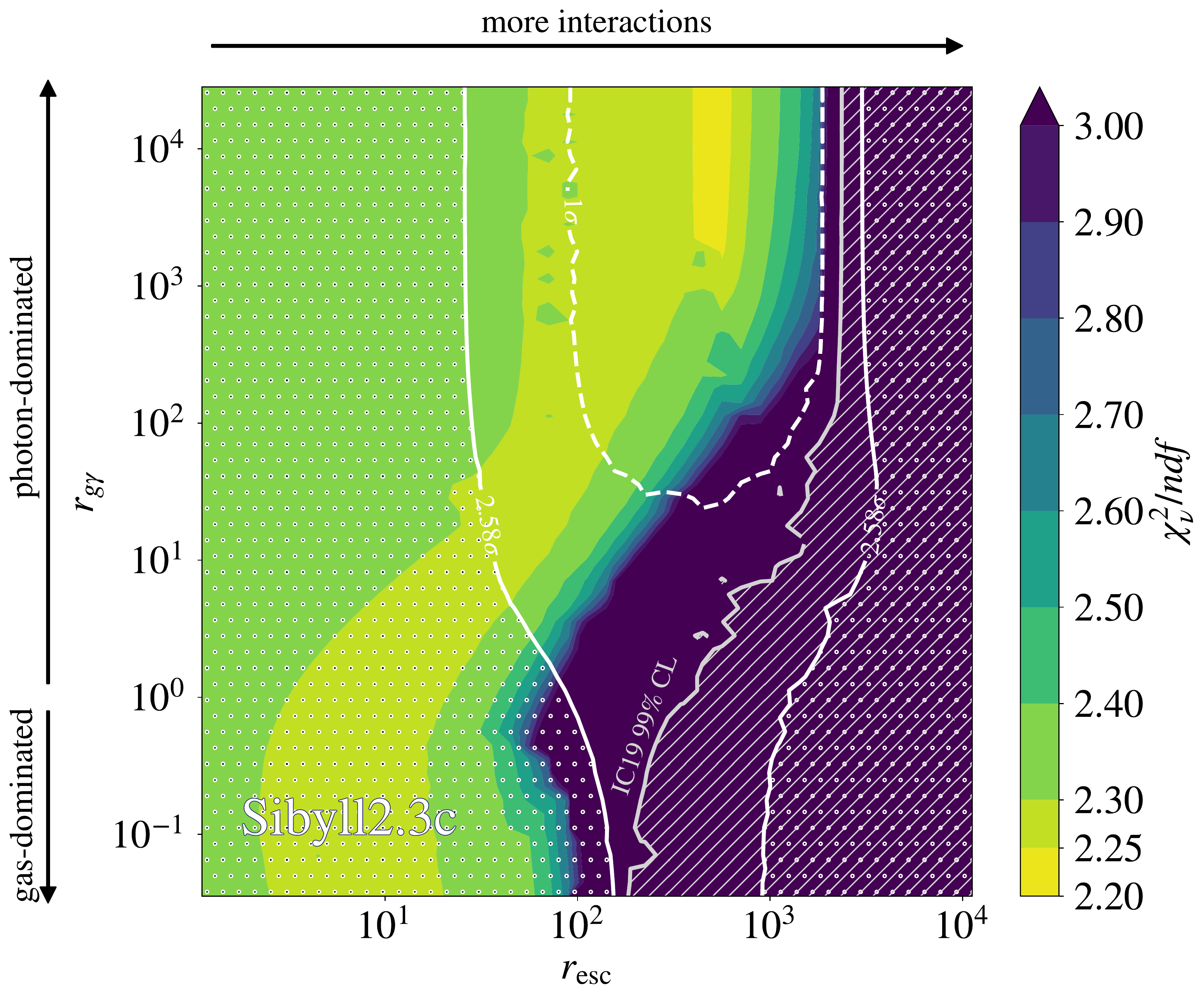}}
    \end{minipage}
    \begin{minipage}{0.49\linewidth}
	  \centering
	  \subfloat[\label{fig:astroChi2_BB_epos}]{\includegraphics[width=\linewidth]{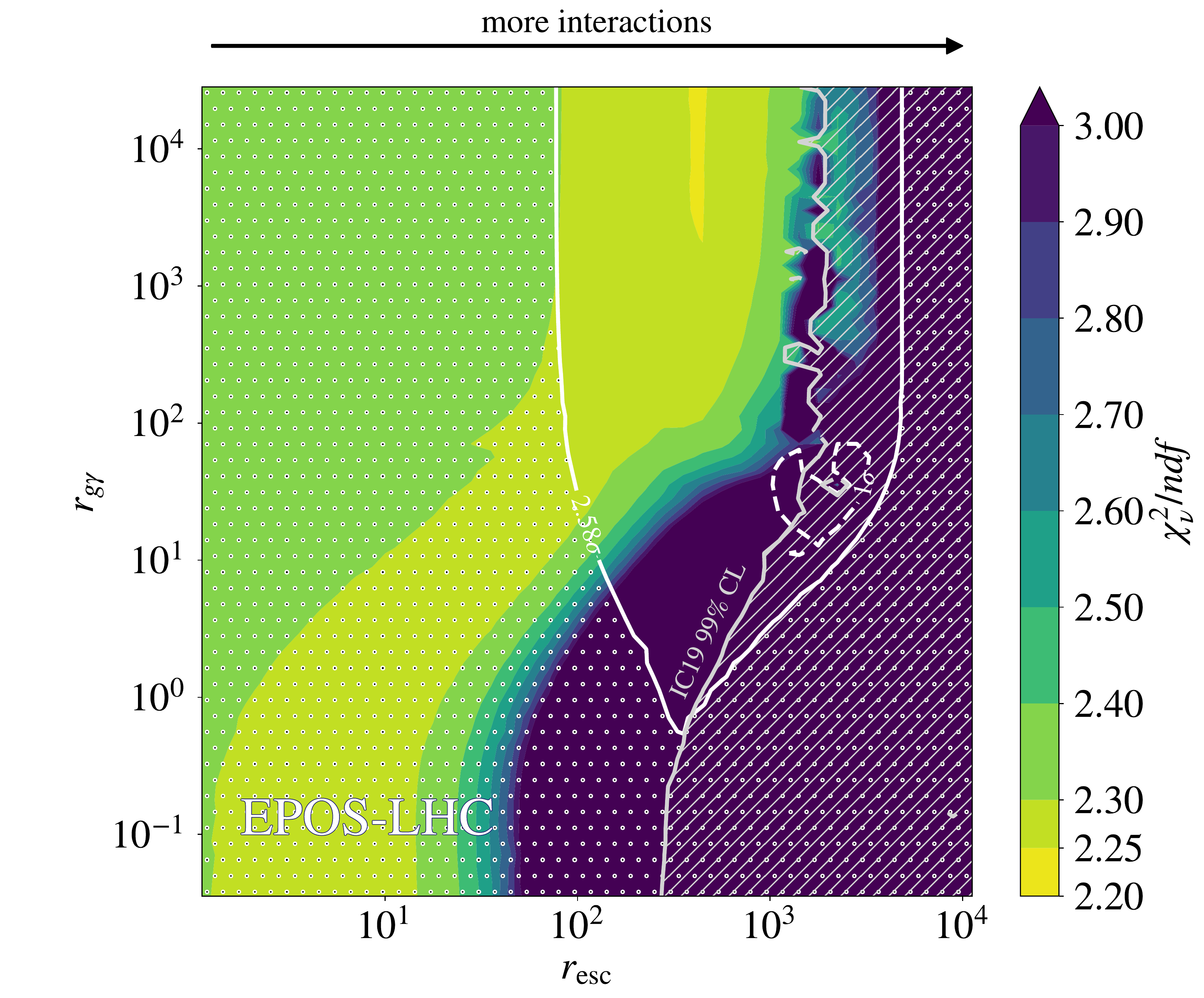}}
    \end{minipage}
    \caption{Reduced $\chi^2_\nu$ of UHECR model predictions of the astrophysical neutrino flux as a function of model parameters $r_\mathrm{esc}\equiv \tau_\mathrm{esc}^\mathrm{ref}/\tau_\mathrm{int}^\mathrm{ref}$ and $r_{g\gamma} \equiv \tau_g^\mathrm{ref} / \tau_\gamma^\mathrm{ref}$ for \textsc{Sibyll2.3c} (left) and \textsc{EPOS-LHC} (right). Contours mark best-fitting region to CR data (dashed white contour) and $99\%$ CL exclusion regions based on UHECR data (solid white contour and dotted region) and non-observation of neutrinos above $10^{15.9}$ eV (grey contour and hatched region).}
    \label{fig:astroChi2_BB}
\end{figure*}

\par
An interesting feature of gas-dominated source environments is that they are able to achieve a good fit to Auger observations even for spectral indices as soft as $\gamma_\mathrm{inj} \sim -2$, as is predicted for diffusive shock acceleration; see Fig. \ref{fig:gammaInj_BB}, which shows the best-fit accelerator spectral index for CRs. This is due to the fact that the hadronic interaction time depends very weakly on energy (see Fig. \ref{fig:timescales}), so that it is well approximated by a single power law $\tau_g \sim E^\zeta$ with $\zeta \sim 0$. Following the analysis of \cite{UFA15} (see Section 2), this implies that quasi-ballistic-regime CRs (i.e. $\tau_\mathrm{esc} \sim E^{-2}$) interacting dominantly with gas, can produce a spectrum escaping the source environment with hard spectral index $\gamma_\mathrm{esc} \gtrsim -1$ even when the injected spectral index is $\gamma_\mathrm{inj} \sim -2$ to $-3$.\footnote{This mechanism can also be achieved by photohadronic interactions, since at high energies $\tau_\gamma \sim E^0$ as well. But in order for the peak of the CR spectrum to be in this regime the photon field needs a very high temperature and CRs require a large number of interactions on average. These two requirements cause the description of the UHECR spectrum and composition to degrade, as well as producing an excess of EHE neutrinos which violates the IceCube bounds.} This clearly demonstrates that processing of the CR spectrum by the source environment can substantially modify the escaping spectral index to be as hard as is required by the Auger spectrum \cite{Aab+16}, without relying on exotic acceleration mechanisms.

\par
Since a soft injected spectral index requires hadronic interactions to dominate to get a good fit to UHECR data, a substantial flux of neutrinos is produced. This creates a potential tension with the IceCube bound on EHE neutrinos, as shown in Fig. \ref{fig:gammaInj_BB}. The viability of a soft injected spectral index will be decisively tested by accurate measurement of the neutrino flux at $\sim 10$ PeV. It is also worth noting that in the gas-dominated regime, where such soft spectral indices can be achieved, \textsc{EPOS-LHC} suffers from tension with both EHE neutrinos and the poor quality of fit it produces to UHECR data. 

\subsection{High energy astrophysical neutrino flux} \label{subsec:hiENus}

\par
The astrophysical neutrino flux has been studied using several different IceCube datasets \cite{IceCubeInelasticity19, IceCubeMuon19, IceCubeCascades20, IceCubeHESE20} with some tension in the various derived neutrino spectra \cite{IceCubeHESE20}. We compare the neutrino flux predicted by our UHECR model to the differential flux measurement presented in the Glashow event paper \cite{IceCubeGlashow2021} and to the IceCube Cascades dataset \cite{IceCubeCascades20}, because this dataset has the least tension with other IceCube measurements of the astrophysical neutrino flux \cite{IceCubeCascades20,IceCubeHESE20}. However, we stress that until the neutrino spectrum is better determined the results of our analysis must be taken as provisional. 

\par
There is no reason to assume that all astrophysical neutrinos are of UHECR origin. Therefore we allow for a component of neutrinos originating from some other source. This non-UHECR component is parametrized by a single power law with an exponential cutoff: $\phi_\nu = \phi_{\nu0} (E/E_0)^{\gamma_\nu} e^{-E/E_{\mathrm{max},\nu}}$, where $E_0 =10^{13}$ eV. Given an UHECR model prediction of the neutrino flux, the normalization, spectral index, and cutoff energy of the non-UHECR neutrinos are tuned so that the total neutrino flux gives the best-fit to the IceCube data.

\par
We measure our model's description of the astrophysical neutrino flux using a $\chi^2$ to the data points,

\begin{align}
  \chi^2 = \displaystyle\sum_i \frac{(\phi_{m,i} - \phi_i)^2}{\sigma_{\phi,i}^2}~. 
\end{align}

\noindent
For energy bins with upper-bounds, we follow the same procedure as for $\chi^2_\mathrm{CR}$ discussed in Section \ref{sec:data}, adding an additional $2n_i$ to the $\chi^2$.
Our final figure of merit is $\chi^2_\nu = \chi^2 + 2\sum_i n_i$, where $i$ runs over energy bins with upper-bounds. We calculate the figure of merit including the Glashow event data point and the portion of the IceCube Cascades dataset within the sensitive energy range, from $16$ TeV to $2.6$ PeV, as determined by IceCube \cite{IceCubeCascades20}. 

\par
Fig. \ref{fig:astroChi2_BB} shows the ability of UHECR models to accurately describe the astrophysical neutrino flux. The best-fitting source environments are photon-dominated with $\langle N_\mathrm{int} \rangle \sim 100-1000$ interactions on average for the reference nucleus. Remarkably, for \textsc{Sibyll2.3c} this region also corresponds to the region best-fitting UHECR data. By contrast these regions are disconnected for \textsc{EPOS-LHC}, making a common origin possible but less natural than for \textsc{Sibyll2.3c}. Given that the region best-fitting the astrophysical neutrino flux corresponds to photon-dominated sources, uncertainties in the modeling of hadronic interactions in the source environment are not relevant to this result.

\par
We note that at low values of $r_\mathrm{esc}$ very few interactions occur before escape so that UHECRs do not produce a substantial neutrino flux. In this case the goodness-of-fit to the astrophysical neutrino flux approaches a constant value because the spectrum is dominated by the non-UHECR component of neutrinos.

\par
\begin{figure*}[htbp!]
	\centering
    \includegraphics[width=\textwidth]{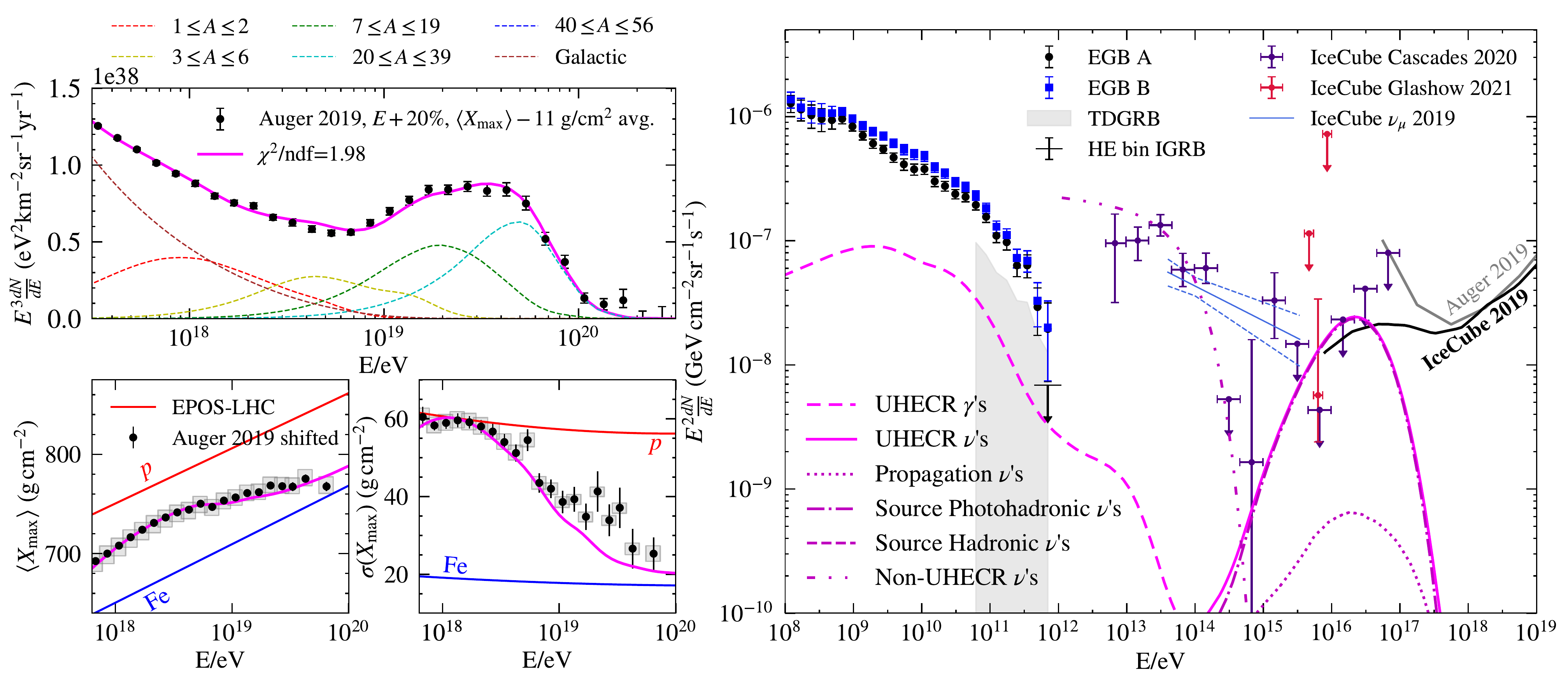}
	\caption{Predictions of the UHECR source model producing the best description of the astrophysical neutrino flux for \textsc{EPOS-LHC}. \textbf{Left:} Same as in Fig. \ref{fig:astroNuBestFits_sibyll}, except the red and blue solid lines show the $\langle X_\mathrm{max}\rangle$ and $\sigma (X_\mathrm{max})$ predictions of \textsc{EPOS-LHC} for pure proton and iron models. \textbf{Right:} Same as in Fig. \ref{fig:astroNuBestFits_sibyll}.}
	\label{fig:astroNuBestFits_epos}
\end{figure*}

The predicted neutrino fluxes with the smallest $\chi^2_\nu$ allowed by multimessenger data are shown in Figs. \ref{fig:astroNuBestFits_sibyll} and \ref{fig:astroNuBestFits_epos} for \textsc{Sibyll2.3c} and \textsc{EPOS-LHC}, respectively.\footnote{We note that our model results in a neutrino spectrum which is photohadronically-produced in the source environment and has a lower peak energy than some other models, e.g. \cite{AlvesBatista+18}, because we account for CR interactions with the source environment's hot (relative to the CMB) photon field. Neutrinos produced during extragalactic UHECR propagation also have a lower peak energy than other models due to the lower average energy-per-nucleon of escaping protons imprinted by interactions in the source environment. These protons primarily produce neutrinos by interacting with the cosmic optical background (COB) which has a peak energy of $\sim 1$ eV.} The parameters for each of these models can be found in Table \ref{tab:astroNuBestFits}. These predictions make clear that UHECR sources may be responsible for the astrophysical neutrino flux at high energies, most importantly producing a sufficient flux to account for the observed Glashow event without conflicting with the bounds on EHE neutrinos. The best-fitting neutrino predictions have a characteristic dip in the neutrino spectrum at $\sim 500$ TeV. This is very similar to the Hypothesis E model investigated by IceCube in \cite{IceCubeCascades20}, which was shown to be favored over a single power-law spectrum by more than $1\sigma$.

\begin{table}[tp]

\centering
\setlength\tabcolsep{7pt}

\begin{tabularx}{\linewidth}{l S S}

\hline \hline
\textbf{Parameter} & \textbf{\textsc{Sibyll2.3c}} & \textbf{\textsc{EPOS-LHC}} \\
\hline
$\gamma_\mathrm{inj}$ & -1.16 & -0.68 \\
$\log_{10}(R_\mathrm{max}/\mathrm{V})$ & 18.59 & 18.47 \\
$\log_{10}{r_\mathrm{esc}}$ & 2.65 & 2.65 \\
$\log_{10}{r_{g\gamma}}$ & 4.35 & 4.45 \\
$\log_{10}(R_\mathrm{diff}/\mathrm{V})$ & 17.79 & 18.14 \\
$\log_{10}{r_\mathrm{size}}$ & 0.98 & 4.93 \\
$T/\mathrm{K}$ & 8987 & 9000 \\
$A_\mathrm{inj}$ & 33.5 & 28.0 \\
$f_\mathrm{gal}$ & 0.82 & 0.79 \\
$\gamma_\mathrm{gal}$ & -3.46 & -3.60 \\
$\log_{10}(E^\mathrm{galFe}_\mathrm{max}/\mathrm{eV})$ & 19.00 & 18.94 \\ 
$A_\mathrm{gal}$ & 26.6 & 30.5 \\
$\phi_{\nu0}$ [$10^{-15}$/(GeV cm$^{2}$ s sr)] & 1.98 & 1.98\\
$\gamma_\nu$ & -2.06 & -2.06 \\
$\log_{10}(E_\mathrm{max,\nu}/\mathrm{eV})$ & 13.89 & 13.89 \\
\hline

\end{tabularx}
\caption{\label{tab:astroNuBestFits} Parameters corresponding to the models presented in Figs. \ref{fig:astroNuBestFits_sibyll} and \ref{fig:astroNuBestFits_epos} for \textsc{Sibyll2.3c} and \textsc{EPOS-LHC}, respectively. $R_\mathrm{max}$ is the maximum rigidity of the injected CR spectrum, where the spectrum is cutoff exponentially; $A_\mathrm{inj}$ is the mass number of the CRs injected into the source environment (non-integers represent the average mass due to a mixture of two consecutive mass numbers); $f_\mathrm{gal}$ is the fraction of the observed flux at $10^{17.55}$ eV which is Galactic; $\gamma_\mathrm{gal}$ is the spectral index, $E^{\gamma_\mathrm{gal}}$, of the Galactic spectrum; $E^\mathrm{galFe}_\mathrm{max}$ is the maximum energy of Galactic iron, where the Galactic component is cutoff exponentially (this parameter sets the maximum rigidity of the Galactic component); $A_\mathrm{gal}$ is the mass number of the Galactic component (this component is also approximated as having a single mass). All other parameters are defined in the text.}

\end{table}

\par
The parameters in Table \ref{tab:astroNuBestFits} show that this description of the astrophysical neutrino flux requires a hard acceleration index $\gamma_\mathrm{inj} \gtrsim -1$ and hot photon field $T \sim 9000$ K (corresponding to a peak photon energy of $\sim 1$ eV), independent of the HIM. In particular, the fit assuming \textsc{EPOS-LHC} additionally relies on a large ratio of source size-to-magnetic coherence length, $r_\mathrm{size} = L/\lambda_c \sim 10^5$, so that CRs at the highest energies enter the quasi-ballistic diffusion regime. Such a large separation of scales may not be possible for real astrophysical sources, so this requirement could be viewed as an additional strain on \textsc{EPOS-LHC} in the case of a common origin scenario. If the common origin scenario could be validated, multimessenger observations would be difficult to reconcile with the source properties required by \textsc{EPOS-LHC}. By contrast, the fit assuming \textsc{Sibyll2.3c} requires only a ratio of source size-to-magnetic coherence length of $\sim 10$. Such a moderate ratio of scales could be accommodated by realistic astrophysical sources and allows \textsc{Sibyll2.3c} to comfortably explain both UHECR data and astrophysical neutrinos. This HIM-dependence of astrophysical parameters is indirect and occurs due to differences in the inferred-composition at Earth. Future work will more systematically investigate the HIM-dependence of astrophysical parameters to determine the robustness of these differences.

\section{Summary}

\par
The main findings of this work can be summarized as follows:
\begin{enumerate}[(i)]
    \item Both photon- and gas-dominated source environments can explain UHECR observations, but gas-dominated environments do so less precisely and are in tension with EHE neutrino constraints (see Fig. \ref{fig:Nsigma_BB}). 
    \item Gas-dominated environments can sufficiently harden CR spectra to fit UHECR data with a soft accelerator spectral index, as in diffusive shock acceleration, if high-rigidity CRs enter the quasi-ballistic diffusion regime before escaping the source environment (see Fig. \ref{fig:gammaInj_BB} and Section \ref{sec:ImpsOfGas}).  
    \item Neutrinos will probe whether UHECR interactions in the source environment are predominantly hadronic or photohadronic, and provide information on the spectral index of the accelerator (see Figs. \ref{fig:Nsigma_BB} and \ref{fig:gammaInj_BB} and Section \ref{sec:ImpsOfGas}). 
    \item UHECR sources may account for high energy astrophysical neutrinos, above $\sim 1$ PeV.  If so, it will be possible constrain hadronic interaction models (see Fig. \ref{fig:astroChi2_BB}, Figs. \ref{fig:astroNuBestFits_sibyll} and \ref{fig:astroNuBestFits_epos}, and Section \ref{subsec:hiENus}).
\end{enumerate}
    
\par
These results directly address the questions presented in Section \ref{sec:intro}. In particular, result (i) directly addresses question (a); results (ii) and (iii) address question (b); and, result (iv) addresses question (c).

\acknowledgments

We would like to acknowledge F. Oikonomou for useful feedback on our analysis and A. Fedynitch, N. Globus, K. Murase, W. Winter, and the anonymous referee for helpful comments on the manuscript. The research of MSM was supported in part by the NYU James Arthur Graduate Award and the Ted Keusseff Fellowship. The research of MSM and GRF was supported in part by NSF-2013199. This work was supported in part through the NYU IT High Performance Computing resources, services, and staff expertise.

\bibliography{MFU21_paper1}

\begin{thebibliography}{51}%
\makeatletter
\providecommand \@ifxundefined [1]{%
 \@ifx{#1\undefined}
}%
\providecommand \@ifnum [1]{%
 \ifnum #1\expandafter \@firstoftwo
 \else \expandafter \@secondoftwo
 \fi
}%
\providecommand \@ifx [1]{%
 \ifx #1\expandafter \@firstoftwo
 \else \expandafter \@secondoftwo
 \fi
}%
\providecommand \natexlab [1]{#1}%
\providecommand \enquote  [1]{``#1''}%
\providecommand \bibnamefont  [1]{#1}%
\providecommand \bibfnamefont [1]{#1}%
\providecommand \citenamefont [1]{#1}%
\providecommand \href@noop [0]{\@secondoftwo}%
\providecommand \href [0]{\begingroup \@sanitize@url \@href}%
\providecommand \@href[1]{\@@startlink{#1}\@@href}%
\providecommand \@@href[1]{\endgroup#1\@@endlink}%
\providecommand \@sanitize@url [0]{\catcode `\\12\catcode `\$12\catcode
  `\&12\catcode `\#12\catcode `\^12\catcode `\_12\catcode `\%12\relax}%
\providecommand \@@startlink[1]{}%
\providecommand \@@endlink[0]{}%
\providecommand \url  [0]{\begingroup\@sanitize@url \@url }%
\providecommand \@url [1]{\endgroup\@href {#1}{\urlprefix }}%
\providecommand \urlprefix  [0]{URL }%
\providecommand \Eprint [0]{\href }%
\providecommand \doibase [0]{https://doi.org/}%
\providecommand \selectlanguage [0]{\@gobble}%
\providecommand \bibinfo  [0]{\@secondoftwo}%
\providecommand \bibfield  [0]{\@secondoftwo}%
\providecommand \translation [1]{[#1]}%
\providecommand \BibitemOpen [0]{}%
\providecommand \bibitemStop [0]{}%
\providecommand \bibitemNoStop [0]{.\EOS\space}%
\providecommand \EOS [0]{\spacefactor3000\relax}%
\providecommand \BibitemShut  [1]{\csname bibitem#1\endcsname}%
\let\auto@bib@innerbib\@empty
\bibitem [{\citenamefont {Anchordoqui}(2019)}]{Anchordoqui18}%
  \BibitemOpen
  \bibfield  {author} {\bibinfo {author} {\bibfnamefont {L.~A.}\ \bibnamefont
  {Anchordoqui}},\ }\href {https://doi.org/10.1016/j.physrep.2019.01.002}
  {\bibfield  {journal} {\bibinfo  {journal} {Phys. Rept.}\ }\textbf {\bibinfo
  {volume} {801}},\ \bibinfo {pages} {1} (\bibinfo {year} {2019})},\ \Eprint
  {https://arxiv.org/abs/1807.09645} {arXiv:1807.09645 [astro-ph.HE]}
  \BibitemShut {NoStop}%
\bibitem [{\citenamefont {{M. G. Aartsen et
  al.}}(2018{\natexlab{a}})}]{TXSObservation}%
  \BibitemOpen
  \bibfield  {author} {\bibinfo {author} {\bibnamefont {{M. G. Aartsen et
  al.}}} (\bibinfo {collaboration} {IceCube, Fermi-LAT, MAGIC, AGILE, ASAS-SN,
  HAWC, H.E.S.S., INTEGRAL, Kanata, Kiso, Kapteyn, Liverpool Telescope, Subaru,
  Swift NuSTAR, VERITAS, VLA/17B-403}),\ }\href
  {https://doi.org/10.1126/science.aat1378} {\bibfield  {journal} {\bibinfo
  {journal} {Science}\ }\textbf {\bibinfo {volume} {361}},\ \bibinfo {pages}
  {eaat1378} (\bibinfo {year} {2018}{\natexlab{a}})},\ \Eprint
  {https://arxiv.org/abs/1807.08816} {arXiv:1807.08816 [astro-ph.HE]}
  \BibitemShut {NoStop}%
\bibitem [{\citenamefont {{R. Stein et al.}}(2021)}]{Stein+20}%
  \BibitemOpen
  \bibfield  {author} {\bibinfo {author} {\bibnamefont {{R. Stein et al.}}},\
  }\href {https://doi.org/10.1038/s41550-020-01295-8} {\bibfield  {journal}
  {\bibinfo  {journal} {Nature Astron.}\ }\textbf {\bibinfo {volume} {5}},\
  \bibinfo {pages} {510} (\bibinfo {year} {2021})},\ \Eprint
  {https://arxiv.org/abs/2005.05340} {arXiv:2005.05340 [astro-ph.HE]}
  \BibitemShut {NoStop}%
\bibitem [{\citenamefont {{M. Unger, G. R. Farrar, and L. A.
  Anchordoqui}}(2015)}]{UFA15}%
  \BibitemOpen
  \bibfield  {author} {\bibinfo {author} {\bibnamefont {{M. Unger, G. R.
  Farrar, and L. A. Anchordoqui}}},\ }\href
  {https://doi.org/10.1103/PhysRevD.92.123001} {\bibfield  {journal} {\bibinfo
  {journal} {Phys. Rev. D}\ }\textbf {\bibinfo {volume} {92}},\ \bibinfo
  {pages} {123001} (\bibinfo {year} {2015})},\ \Eprint
  {https://arxiv.org/abs/1505.02153} {arXiv:1505.02153 [astro-ph.HE]}
  \BibitemShut {NoStop}%
\bibitem [{\citenamefont {{M. S. Muzio, M. Unger, and G. R.
  Farrar}}(2019)}]{MUF19}%
  \BibitemOpen
  \bibfield  {author} {\bibinfo {author} {\bibnamefont {{M. S. Muzio, M. Unger,
  and G. R. Farrar}}},\ }\href {https://doi.org/10.1103/PhysRevD.100.103008}
  {\bibfield  {journal} {\bibinfo  {journal} {Phys. Rev. D}\ }\textbf {\bibinfo
  {volume} {100}},\ \bibinfo {pages} {103008} (\bibinfo {year} {2019})},\
  \Eprint {https://arxiv.org/abs/1906.06233} {arXiv:1906.06233 [astro-ph.HE]}
  \BibitemShut {NoStop}%
\bibitem [{\citenamefont {{G. Giacinti, M. Kachelrie\ss{}, O. Kalashev, A.
  Neronov, and D. V. Semikoz}}(2015)}]{Giacinti+15}%
  \BibitemOpen
  \bibfield  {author} {\bibinfo {author} {\bibnamefont {{G. Giacinti, M.
  Kachelrie\ss{}, O. Kalashev, A. Neronov, and D. V. Semikoz}}},\ }\href
  {https://doi.org/10.1103/PhysRevD.92.083016} {\bibfield  {journal} {\bibinfo
  {journal} {Phys. Rev. D}\ }\textbf {\bibinfo {volume} {92}},\ \bibinfo
  {pages} {083016} (\bibinfo {year} {2015})},\ \Eprint
  {https://arxiv.org/abs/1507.07534} {arXiv:1507.07534 [astro-ph.HE]}
  \BibitemShut {NoStop}%
\bibitem [{\citenamefont {{M. Kachelrie{\ss} et al.}}(2017)}]{Kachelriess+17}%
  \BibitemOpen
  \bibfield  {author} {\bibinfo {author} {\bibnamefont {{M. Kachelrie{\ss} et
  al.}}},\ }\href {https://doi.org/10.1103/PhysRevD.96.083006} {\bibfield
  {journal} {\bibinfo  {journal} {Phys. Rev. D}\ }\textbf {\bibinfo {volume}
  {96}},\ \bibinfo {pages} {083006} (\bibinfo {year} {2017})},\ \Eprint
  {https://arxiv.org/abs/1704.06893} {arXiv:1704.06893 [astro-ph.HE]}
  \BibitemShut {NoStop}%
\bibitem [{\citenamefont {{S. Yoshida and K. Murase}}(2020)}]{Yoshida+20}%
  \BibitemOpen
  \bibfield  {author} {\bibinfo {author} {\bibnamefont {{S. Yoshida and K.
  Murase}}},\ }\href {https://doi.org/10.1103/PhysRevD.102.083023} {\bibfield
  {journal} {\bibinfo  {journal} {Phys. Rev. D}\ }\textbf {\bibinfo {volume}
  {102}},\ \bibinfo {pages} {083023} (\bibinfo {year} {2020})},\ \Eprint
  {https://arxiv.org/abs/2007.09276} {arXiv:2007.09276 [astro-ph.HE]}
  \BibitemShut {NoStop}%
\bibitem [{\citenamefont {{D. Biehl, D. Boncioli, and C. Lunardini, and W.
  Winter}}(2018)}]{Biehl+17b}%
  \BibitemOpen
  \bibfield  {author} {\bibinfo {author} {\bibnamefont {{D. Biehl, D. Boncioli,
  and C. Lunardini, and W. Winter}}},\ }\href
  {https://doi.org/10.1038/s41598-018-29022-4} {\bibfield  {journal} {\bibinfo
  {journal} {Sci. Rep.}\ }\textbf {\bibinfo {volume} {8}},\ \bibinfo {pages}
  {10828} (\bibinfo {year} {2018})},\ \Eprint
  {https://arxiv.org/abs/1711.03555} {arXiv:1711.03555 [astro-ph.HE]}
  \BibitemShut {NoStop}%
\bibitem [{\citenamefont {Fang}\ and\ \citenamefont {Murase}(2018)}]{Fang+17}%
  \BibitemOpen
  \bibfield  {author} {\bibinfo {author} {\bibfnamefont {K.}~\bibnamefont
  {Fang}}\ and\ \bibinfo {author} {\bibfnamefont {K.}~\bibnamefont {Murase}},\
  }\href {https://doi.org/10.1038/s41567-017-0025-4} {\bibfield  {journal}
  {\bibinfo  {journal} {Nature Phys.}\ }\textbf {\bibinfo {volume} {14}},\
  \bibinfo {pages} {396} (\bibinfo {year} {2018})},\ \Eprint
  {https://arxiv.org/abs/1704.00015} {arXiv:1704.00015 [astro-ph.HE]}
  \BibitemShut {NoStop}%
\bibitem [{\citenamefont {{D. Boncioli, D. Biehl, Daniel and W.
  Winter}}(2019)}]{Boncioli+18}%
  \BibitemOpen
  \bibfield  {author} {\bibinfo {author} {\bibnamefont {{D. Boncioli, D. Biehl,
  Daniel and W. Winter}}},\ }\href {https://doi.org/10.3847/1538-4357/aafda7}
  {\bibfield  {journal} {\bibinfo  {journal} {Astrophys. J.}\ }\textbf
  {\bibinfo {volume} {872}},\ \bibinfo {pages} {110} (\bibinfo {year}
  {2019})},\ \Eprint {https://arxiv.org/abs/1808.07481} {arXiv:1808.07481
  [astro-ph.HE]} \BibitemShut {NoStop}%
\bibitem [{\citenamefont {Gu\'epin}\ \emph {et~al.}(2018)\citenamefont
  {Gu\'epin}, \citenamefont {Kotera}, \citenamefont {Barausse}, \citenamefont
  {Fang},\ and\ \citenamefont {Murase}}]{Guepin+17}%
  \BibitemOpen
  \bibfield  {author} {\bibinfo {author} {\bibfnamefont {C.}~\bibnamefont
  {Gu\'epin}}, \bibinfo {author} {\bibfnamefont {K.}~\bibnamefont {Kotera}},
  \bibinfo {author} {\bibfnamefont {E.}~\bibnamefont {Barausse}}, \bibinfo
  {author} {\bibfnamefont {K.}~\bibnamefont {Fang}},\ and\ \bibinfo {author}
  {\bibfnamefont {K.}~\bibnamefont {Murase}},\ }\href
  {https://doi.org/10.1051/0004-6361/201732392} {\bibfield  {journal} {\bibinfo
   {journal} {Astron. Astrophys.}\ }\textbf {\bibinfo {volume} {616}},\
  \bibinfo {pages} {A179} (\bibinfo {year} {2018})},\ \bibinfo {note}
  {[Erratum: Astron.Astrophys. 636, C3 (2020)]},\ \Eprint
  {https://arxiv.org/abs/1711.11274} {arXiv:1711.11274 [astro-ph.HE]}
  \BibitemShut {NoStop}%
\bibitem [{\citenamefont {Zhang}\ and\ \citenamefont
  {Murase}(2019)}]{Zhang+18}%
  \BibitemOpen
  \bibfield  {author} {\bibinfo {author} {\bibfnamefont {B.~T.}\ \bibnamefont
  {Zhang}}\ and\ \bibinfo {author} {\bibfnamefont {K.}~\bibnamefont {Murase}},\
  }\href {https://doi.org/10.1103/PhysRevD.100.103004} {\bibfield  {journal}
  {\bibinfo  {journal} {Phys. Rev. D}\ }\textbf {\bibinfo {volume} {100}},\
  \bibinfo {pages} {103004} (\bibinfo {year} {2019})},\ \Eprint
  {https://arxiv.org/abs/1812.10289} {arXiv:1812.10289 [astro-ph.HE]}
  \BibitemShut {NoStop}%
\bibitem [{\citenamefont {{X. Rodrigues et al.}}(2021)}]{Rodrigues+20}%
  \BibitemOpen
  \bibfield  {author} {\bibinfo {author} {\bibnamefont {{X. Rodrigues et
  al.}}},\ }\href {https://doi.org/10.1103/PhysRevLett.126.191101} {\bibfield
  {journal} {\bibinfo  {journal} {Phys. Rev. Lett.}\ }\textbf {\bibinfo
  {volume} {126}},\ \bibinfo {pages} {191101} (\bibinfo {year} {2021})},\
  \Eprint {https://arxiv.org/abs/2003.08392} {arXiv:2003.08392 [astro-ph.HE]}
  \BibitemShut {NoStop}%
\bibitem [{\citenamefont {{D. Biehl et al.}}(2018)}]{Biehl+17a}%
  \BibitemOpen
  \bibfield  {author} {\bibinfo {author} {\bibnamefont {{D. Biehl et al.}}},\
  }\href {https://doi.org/10.1051/0004-6361/201731337} {\bibfield  {journal}
  {\bibinfo  {journal} {Astron. Astrophys.}\ }\textbf {\bibinfo {volume}
  {611}},\ \bibinfo {pages} {A101} (\bibinfo {year} {2018})},\ \Eprint
  {https://arxiv.org/abs/1705.08909} {arXiv:1705.08909 [astro-ph.HE]}
  \BibitemShut {NoStop}%
\bibitem [{\citenamefont {Globus}\ \emph {et~al.}(2017)\citenamefont {Globus},
  \citenamefont {Allard}, \citenamefont {Parizot},\ and\ \citenamefont
  {Piran}}]{Globus+17}%
  \BibitemOpen
  \bibfield  {author} {\bibinfo {author} {\bibfnamefont {N.}~\bibnamefont
  {Globus}}, \bibinfo {author} {\bibfnamefont {D.}~\bibnamefont {Allard}},
  \bibinfo {author} {\bibfnamefont {E.}~\bibnamefont {Parizot}},\ and\ \bibinfo
  {author} {\bibfnamefont {T.}~\bibnamefont {Piran}},\ }\href
  {https://doi.org/10.3847/2041-8213/aa6af0} {\bibfield  {journal} {\bibinfo
  {journal} {Astrophys. J. Lett.}\ }\textbf {\bibinfo {volume} {839}},\
  \bibinfo {pages} {L22} (\bibinfo {year} {2017})},\ \Eprint
  {https://arxiv.org/abs/1703.04158} {arXiv:1703.04158 [astro-ph.HE]}
  \BibitemShut {NoStop}%
\bibitem [{\citenamefont {Alves~Batista}\ \emph {et~al.}(2019)\citenamefont
  {Alves~Batista}, \citenamefont {de~Almeida}, \citenamefont {Lago},\ and\
  \citenamefont {Kotera}}]{AlvesBatista+18}%
  \BibitemOpen
  \bibfield  {author} {\bibinfo {author} {\bibfnamefont {R.}~\bibnamefont
  {Alves~Batista}}, \bibinfo {author} {\bibfnamefont {R.~M.}\ \bibnamefont
  {de~Almeida}}, \bibinfo {author} {\bibfnamefont {B.}~\bibnamefont {Lago}},\
  and\ \bibinfo {author} {\bibfnamefont {K.}~\bibnamefont {Kotera}},\ }\href
  {https://doi.org/10.1088/1475-7516/2019/01/002} {\bibfield  {journal}
  {\bibinfo  {journal} {JCAP}\ }\textbf {\bibinfo {volume} {01}},\ \bibinfo
  {pages} {002}},\ \Eprint {https://arxiv.org/abs/1806.10879} {arXiv:1806.10879
  [astro-ph.HE]} \BibitemShut {NoStop}%
\bibitem [{\citenamefont {Heinze}\ \emph {et~al.}(2020)\citenamefont {Heinze},
  \citenamefont {Biehl}, \citenamefont {Fedynitch}, \citenamefont {Boncioli},
  \citenamefont {Rudolph},\ and\ \citenamefont {Winter}}]{Heinze+20}%
  \BibitemOpen
  \bibfield  {author} {\bibinfo {author} {\bibfnamefont {J.}~\bibnamefont
  {Heinze}}, \bibinfo {author} {\bibfnamefont {D.}~\bibnamefont {Biehl}},
  \bibinfo {author} {\bibfnamefont {A.}~\bibnamefont {Fedynitch}}, \bibinfo
  {author} {\bibfnamefont {D.}~\bibnamefont {Boncioli}}, \bibinfo {author}
  {\bibfnamefont {A.}~\bibnamefont {Rudolph}},\ and\ \bibinfo {author}
  {\bibfnamefont {W.}~\bibnamefont {Winter}},\ }\href
  {https://doi.org/10.1093/mnras/staa2751} {\bibfield  {journal} {\bibinfo
  {journal} {Mon. Not. Roy. Astron. Soc.}\ }\textbf {\bibinfo {volume} {498}},\
  \bibinfo {pages} {5990} (\bibinfo {year} {2020})},\ \Eprint
  {https://arxiv.org/abs/2006.14301} {arXiv:2006.14301 [astro-ph.HE]}
  \BibitemShut {NoStop}%
\bibitem [{\citenamefont {{E. Waxman and J. N. Bahcall}}(1999)}]{Waxman+98}%
  \BibitemOpen
  \bibfield  {author} {\bibinfo {author} {\bibnamefont {{E. Waxman and J. N.
  Bahcall}}},\ }\href {https://doi.org/10.1103/PhysRevD.59.023002} {\bibfield
  {journal} {\bibinfo  {journal} {Phys. Rev. D}\ }\textbf {\bibinfo {volume}
  {59}},\ \bibinfo {pages} {023002} (\bibinfo {year} {1999})},\ \Eprint
  {https://arxiv.org/abs/hep-ph/9807282} {arXiv:hep-ph/9807282} \BibitemShut
  {NoStop}%
\bibitem [{\citenamefont {{A. Aab et al.}}(2020{\natexlab{a}})}]{Aab+20a}%
  \BibitemOpen
  \bibfield  {author} {\bibinfo {author} {\bibnamefont {{A. Aab et al.}}}
  (\bibinfo {collaboration} {Pierre Auger}),\ }\href
  {https://doi.org/10.1103/PhysRevD.102.062005} {\bibfield  {journal} {\bibinfo
   {journal} {Phys. Rev. D}\ }\textbf {\bibinfo {volume} {102}},\ \bibinfo
  {pages} {062005} (\bibinfo {year} {2020}{\natexlab{a}})},\ \Eprint
  {https://arxiv.org/abs/2008.06486} {arXiv:2008.06486 [astro-ph.HE]}
  \BibitemShut {NoStop}%
\bibitem [{\citenamefont {{A. Aab et al.}}(2020{\natexlab{b}})}]{Aab+20b}%
  \BibitemOpen
  \bibfield  {author} {\bibinfo {author} {\bibnamefont {{A. Aab et al.}}}
  (\bibinfo {collaboration} {Pierre Auger}),\ }\href
  {https://doi.org/10.1103/PhysRevLett.125.121106} {\bibfield  {journal}
  {\bibinfo  {journal} {Phys. Rev. Lett.}\ }\textbf {\bibinfo {volume} {125}},\
  \bibinfo {pages} {121106} (\bibinfo {year} {2020}{\natexlab{b}})},\ \Eprint
  {https://arxiv.org/abs/2008.06488} {arXiv:2008.06488 [astro-ph.HE]}
  \BibitemShut {NoStop}%
\bibitem [{\citenamefont {Verzi}(2020)}]{Verzi20}%
  \BibitemOpen
  \bibfield  {author} {\bibinfo {author} {\bibfnamefont {V.}~\bibnamefont
  {Verzi}} (\bibinfo {collaboration} {Pierre Auger}),\ }\href
  {https://doi.org/10.22323/1.358.0450} {\bibfield  {journal} {\bibinfo
  {journal} {PoS}\ }\textbf {\bibinfo {volume} {ICRC2019}},\ \bibinfo {pages}
  {450} (\bibinfo {year} {2020})}\BibitemShut {NoStop}%
\bibitem [{\citenamefont {Yushkov}(2020)}]{Yushkov20}%
  \BibitemOpen
  \bibfield  {author} {\bibinfo {author} {\bibfnamefont {A.}~\bibnamefont
  {Yushkov}} (\bibinfo {collaboration} {Auger}),\ }\href
  {https://doi.org/10.22323/1.358.0482} {\bibfield  {journal} {\bibinfo
  {journal} {PoS}\ }\textbf {\bibinfo {volume} {ICRC2019}},\ \bibinfo {pages}
  {482} (\bibinfo {year} {2020})}\BibitemShut {NoStop}%
\bibitem [{\citenamefont {{M. Ackermann et al.}}(2015)}]{Ackermann+14}%
  \BibitemOpen
  \bibfield  {author} {\bibinfo {author} {\bibnamefont {{M. Ackermann et al.}}}
  (\bibinfo {collaboration} {Fermi-LAT}),\ }\href
  {https://doi.org/10.1088/0004-637X/799/1/86} {\bibfield  {journal} {\bibinfo
  {journal} {Astrophys. J.}\ }\textbf {\bibinfo {volume} {799}},\ \bibinfo
  {pages} {86} (\bibinfo {year} {2015})},\ \Eprint
  {https://arxiv.org/abs/1410.3696} {arXiv:1410.3696 [astro-ph.HE]}
  \BibitemShut {NoStop}%
\bibitem [{\citenamefont {{M. G. Aartsen et al.}}(2020)}]{IceCubeCascades20}%
  \BibitemOpen
  \bibfield  {author} {\bibinfo {author} {\bibnamefont {{M. G. Aartsen et
  al.}}} (\bibinfo {collaboration} {IceCube}),\ }\href
  {https://doi.org/10.1103/PhysRevLett.125.121104} {\bibfield  {journal}
  {\bibinfo  {journal} {Phys. Rev. Lett.}\ }\textbf {\bibinfo {volume} {125}},\
  \bibinfo {pages} {121104} (\bibinfo {year} {2020})},\ \Eprint
  {https://arxiv.org/abs/2001.09520} {arXiv:2001.09520 [astro-ph.HE]}
  \BibitemShut {NoStop}%
\bibitem [{\citenamefont {Stettner}(2020)}]{IceCubeMuon19}%
  \BibitemOpen
  \bibfield  {author} {\bibinfo {author} {\bibfnamefont {J.}~\bibnamefont
  {Stettner}} (\bibinfo {collaboration} {IceCube}),\ }\href
  {https://doi.org/10.22323/1.358.1017} {\bibfield  {journal} {\bibinfo
  {journal} {PoS}\ }\textbf {\bibinfo {volume} {ICRC2019}},\ \bibinfo {pages}
  {1017} (\bibinfo {year} {2020})},\ \Eprint {https://arxiv.org/abs/1908.09551}
  {arXiv:1908.09551 [astro-ph.HE]} \BibitemShut {NoStop}%
\bibitem [{\citenamefont {{M. G. Aartsen, et al.}}(2021)}]{IceCubeGlashow2021}%
  \BibitemOpen
  \bibfield  {author} {\bibinfo {author} {\bibnamefont {{M. G. Aartsen, et
  al.}}} (\bibinfo {collaboration} {IceCube}),\ }\href
  {https://doi.org/10.1038/s41586-021-03256-1} {\bibfield  {journal} {\bibinfo
  {journal} {Nature}\ }\textbf {\bibinfo {volume} {591}},\ \bibinfo {pages}
  {220} (\bibinfo {year} {2021})}\BibitemShut {NoStop}%
\bibitem [{\citenamefont {{M. G. Aartsen et
  al.}}(2018{\natexlab{b}})}]{Aartsen+18}%
  \BibitemOpen
  \bibfield  {author} {\bibinfo {author} {\bibnamefont {{M. G. Aartsen et
  al.}}} (\bibinfo {collaboration} {IceCube}),\ }\href
  {https://doi.org/10.1103/PhysRevD.98.062003} {\bibfield  {journal} {\bibinfo
  {journal} {Phys. Rev. D}\ }\textbf {\bibinfo {volume} {98}},\ \bibinfo
  {pages} {062003} (\bibinfo {year} {2018}{\natexlab{b}})},\ \Eprint
  {https://arxiv.org/abs/1807.01820} {arXiv:1807.01820 [astro-ph.HE]}
  \BibitemShut {NoStop}%
\bibitem [{\citenamefont {{A. Aab et al.}}(2019)}]{Aab+2019}%
  \BibitemOpen
  \bibfield  {author} {\bibinfo {author} {\bibnamefont {{A. Aab et al.}}}
  (\bibinfo {collaboration} {Pierre Auger}),\ }\href
  {https://doi.org/10.1088/1475-7516/2019/10/022} {\bibfield  {journal}
  {\bibinfo  {journal} {JCAP}\ }\textbf {\bibinfo {volume} {10}},\ \bibinfo
  {pages} {022}},\ \Eprint {https://arxiv.org/abs/1906.07422} {arXiv:1906.07422
  [astro-ph.HE]} \BibitemShut {NoStop}%
\bibitem [{\citenamefont {{B. Robertson, R. Ellis, S. Furlanetto, and J.
  Dunlop}}(2015)}]{Robertson+15}%
  \BibitemOpen
  \bibfield  {author} {\bibinfo {author} {\bibnamefont {{B. Robertson, R.
  Ellis, S. Furlanetto, and J. Dunlop}}},\ }\href
  {https://doi.org/10.1088/2041-8205/802/2/L19} {\bibfield  {journal} {\bibinfo
   {journal} {Astrophys. J.}\ }\textbf {\bibinfo {volume} {802}},\ \bibinfo
  {pages} {19} (\bibinfo {year} {2015})},\ \Eprint
  {https://arxiv.org/abs/1502.02024} {arXiv:1502.02024 [astro-ph.CO]}
  \BibitemShut {NoStop}%
\bibitem [{\citenamefont {{Aab, Alexander et al.}}(2017)}]{Aab+16}%
  \BibitemOpen
  \bibfield  {author} {\bibinfo {author} {\bibnamefont {{Aab, Alexander et
  al.}}} (\bibinfo {collaboration} {Pierre Auger}),\ }\href
  {https://doi.org/10.1088/1475-7516/2018/03/E02,
  10.1088/1475-7516/2017/04/038} {\bibfield  {journal} {\bibinfo  {journal}
  {JCAP}\ }\textbf {\bibinfo {volume} {1704}}\bibfield  {number} {\bibinfo
  {number} { (04)},\ \bibinfo {pages} {038}},\ }\bibinfo {note} {[Erratum:
  JCAP1803,no.03,E02(2018)]},\ \Eprint {https://arxiv.org/abs/1612.07155}
  {arXiv:1612.07155 [astro-ph.HE]} \BibitemShut {NoStop}%
\bibitem [{\citenamefont {{N. Globus, D. Allard, and E.
  Parizot}}(2008)}]{Globus+07}%
  \BibitemOpen
  \bibfield  {author} {\bibinfo {author} {\bibnamefont {{N. Globus, D. Allard,
  and E. Parizot}}},\ }\href {https://doi.org/10.1051/0004-6361:20078653}
  {\bibfield  {journal} {\bibinfo  {journal} {Astron. Astrophys.}\ }\textbf
  {\bibinfo {volume} {479}},\ \bibinfo {pages} {97} (\bibinfo {year} {2008})},\
  \Eprint {https://arxiv.org/abs/0709.1541} {arXiv:0709.1541 [astro-ph]}
  \BibitemShut {NoStop}%
\bibitem [{\citenamefont {{S. Mollerach and E. Roulet}}(2013)}]{Mollerach+13}%
  \BibitemOpen
  \bibfield  {author} {\bibinfo {author} {\bibnamefont {{S. Mollerach and E.
  Roulet}}},\ }\href {https://doi.org/10.1088/1475-7516/2013/10/013} {\bibfield
   {journal} {\bibinfo  {journal} {JCAP}\ }\textbf {\bibinfo {volume} {10}},\
  \bibinfo {pages} {013}},\ \Eprint {https://arxiv.org/abs/1305.6519}
  {arXiv:1305.6519 [astro-ph.HE]} \BibitemShut {NoStop}%
\bibitem [{Wit(2017)}]{Wittkowski17}%
  \BibitemOpen
  \href@noop {} {\emph {\bibinfo {title} {{The Pierre Auger Observatory:
  Contributions to the 35th International Cosmic Ray Conference (ICRC
  2017)}}}}\ (\bibinfo {year} {2017})\ \Eprint
  {https://arxiv.org/abs/1708.06592} {arXiv:1708.06592 [astro-ph.HE]}
  \BibitemShut {NoStop}%
\bibitem [{\citenamefont {Fiorillo}\ \emph {et~al.}(2021)\citenamefont
  {Fiorillo}, \citenamefont {Van~Vliet}, \citenamefont {Morisi},\ and\
  \citenamefont {Winter}}]{Fiorillo+21}%
  \BibitemOpen
  \bibfield  {author} {\bibinfo {author} {\bibfnamefont {D.~F.~G.}\
  \bibnamefont {Fiorillo}}, \bibinfo {author} {\bibfnamefont {A.}~\bibnamefont
  {Van~Vliet}}, \bibinfo {author} {\bibfnamefont {S.}~\bibnamefont {Morisi}},\
  and\ \bibinfo {author} {\bibfnamefont {W.}~\bibnamefont {Winter}},\ }\href
  {https://doi.org/10.1088/1475-7516/2021/07/028} {\bibfield  {journal}
  {\bibinfo  {journal} {JCAP}\ }\textbf {\bibinfo {volume} {07}},\ \bibinfo
  {pages} {028}},\ \Eprint {https://arxiv.org/abs/2103.16577} {arXiv:2103.16577
  [astro-ph.HE]} \BibitemShut {NoStop}%
\bibitem [{\citenamefont {Alves~Batista}\ \emph {et~al.}(2016)\citenamefont
  {Alves~Batista}, \citenamefont {Dundovic}, \citenamefont {Erdmann},
  \citenamefont {Kampert}, \citenamefont {Kuempel}, \citenamefont {M\"uller},
  \citenamefont {Sigl}, \citenamefont {van Vliet}, \citenamefont {Walz},\ and\
  \citenamefont {Winchen}}]{AlvesBatista+16}%
  \BibitemOpen
  \bibfield  {author} {\bibinfo {author} {\bibfnamefont {R.}~\bibnamefont
  {Alves~Batista}}, \bibinfo {author} {\bibfnamefont {A.}~\bibnamefont
  {Dundovic}}, \bibinfo {author} {\bibfnamefont {M.}~\bibnamefont {Erdmann}},
  \bibinfo {author} {\bibfnamefont {K.-H.}\ \bibnamefont {Kampert}}, \bibinfo
  {author} {\bibfnamefont {D.}~\bibnamefont {Kuempel}}, \bibinfo {author}
  {\bibfnamefont {G.}~\bibnamefont {M\"uller}}, \bibinfo {author}
  {\bibfnamefont {G.}~\bibnamefont {Sigl}}, \bibinfo {author} {\bibfnamefont
  {A.}~\bibnamefont {van Vliet}}, \bibinfo {author} {\bibfnamefont
  {D.}~\bibnamefont {Walz}},\ and\ \bibinfo {author} {\bibfnamefont
  {T.}~\bibnamefont {Winchen}},\ }\href
  {https://doi.org/10.1088/1475-7516/2016/05/038} {\bibfield  {journal}
  {\bibinfo  {journal} {JCAP}\ }\textbf {\bibinfo {volume} {05}},\ \bibinfo
  {pages} {038}},\ \Eprint {https://arxiv.org/abs/1603.07142} {arXiv:1603.07142
  [astro-ph.IM]} \BibitemShut {NoStop}%
\bibitem [{\citenamefont {{C. Baus, T. Pierog, and R. Ulrich}}(2016)}]{CRMC}%
  \BibitemOpen
  \bibfield  {author} {\bibinfo {author} {\bibnamefont {{C. Baus, T. Pierog,
  and R. Ulrich}}},\ }\href@noop {} {\bibinfo {title} {{CRMC (Cosmic Ray Monte
  Carlo package)}}},\ \bibinfo {howpublished}
  {\url{https://github.com/alisw/crmc}} (\bibinfo {year} {2016})\BibitemShut
  {NoStop}%
\bibitem [{\citenamefont {{T. Pierog et al.}}(2015)}]{Pierog+13}%
  \BibitemOpen
  \bibfield  {author} {\bibinfo {author} {\bibnamefont {{T. Pierog et al.}}},\
  }\href {https://doi.org/10.1103/PhysRevC.92.034906} {\bibfield  {journal}
  {\bibinfo  {journal} {Phys. Rev.}\ }\textbf {\bibinfo {volume} {C92}},\
  \bibinfo {pages} {034906} (\bibinfo {year} {2015})},\ \Eprint
  {https://arxiv.org/abs/1306.0121} {arXiv:1306.0121 [hep-ph]} \BibitemShut
  {NoStop}%
\bibitem [{\citenamefont {Fedynitch}\ \emph {et~al.}(2019)\citenamefont
  {Fedynitch}, \citenamefont {Riehn}, \citenamefont {Engel}, \citenamefont
  {Gaisser},\ and\ \citenamefont {Stanev}}]{Fedynitch+18}%
  \BibitemOpen
  \bibfield  {author} {\bibinfo {author} {\bibfnamefont {A.}~\bibnamefont
  {Fedynitch}}, \bibinfo {author} {\bibfnamefont {F.}~\bibnamefont {Riehn}},
  \bibinfo {author} {\bibfnamefont {R.}~\bibnamefont {Engel}}, \bibinfo
  {author} {\bibfnamefont {T.~K.}\ \bibnamefont {Gaisser}},\ and\ \bibinfo
  {author} {\bibfnamefont {T.}~\bibnamefont {Stanev}},\ }\href
  {https://doi.org/10.1103/PhysRevD.100.103018} {\bibfield  {journal} {\bibinfo
   {journal} {Phys. Rev. D}\ }\textbf {\bibinfo {volume} {100}},\ \bibinfo
  {pages} {103018} (\bibinfo {year} {2019})},\ \Eprint
  {https://arxiv.org/abs/1806.04140} {arXiv:1806.04140 [hep-ph]} \BibitemShut
  {NoStop}%
\bibitem [{\citenamefont {Baerwald}\ \emph {et~al.}(2012)\citenamefont
  {Baerwald}, \citenamefont {Hummer},\ and\ \citenamefont
  {Winter}}]{Baerwald+11}%
  \BibitemOpen
  \bibfield  {author} {\bibinfo {author} {\bibfnamefont {P.}~\bibnamefont
  {Baerwald}}, \bibinfo {author} {\bibfnamefont {S.}~\bibnamefont {Hummer}},\
  and\ \bibinfo {author} {\bibfnamefont {W.}~\bibnamefont {Winter}},\ }\href
  {https://doi.org/10.1016/j.astropartphys.2011.11.005} {\bibfield  {journal}
  {\bibinfo  {journal} {Astropart. Phys.}\ }\textbf {\bibinfo {volume} {35}},\
  \bibinfo {pages} {508} (\bibinfo {year} {2012})},\ \Eprint
  {https://arxiv.org/abs/1107.5583} {arXiv:1107.5583 [astro-ph.HE]}
  \BibitemShut {NoStop}%
\bibitem [{\citenamefont {{D. Harari, S. Mollerach, and E.
  Roulet}}(2014)}]{Harari+13}%
  \BibitemOpen
  \bibfield  {author} {\bibinfo {author} {\bibnamefont {{D. Harari, S.
  Mollerach, and E. Roulet}}},\ }\href
  {https://doi.org/10.1103/PhysRevD.89.123001} {\bibfield  {journal} {\bibinfo
  {journal} {Phys. Rev. D}\ }\textbf {\bibinfo {volume} {89}},\ \bibinfo
  {pages} {123001} (\bibinfo {year} {2014})},\ \Eprint
  {https://arxiv.org/abs/1312.1366} {arXiv:1312.1366 [astro-ph.HE]}
  \BibitemShut {NoStop}%
\bibitem [{\citenamefont {{P. Abreu et al.}}(2013)}]{Abreu+13}%
  \BibitemOpen
  \bibfield  {author} {\bibinfo {author} {\bibnamefont {{P. Abreu et al.}}}
  (\bibinfo {collaboration} {Pierre Auger}),\ }\href
  {https://doi.org/10.1088/1475-7516/2013/02/026} {\bibfield  {journal}
  {\bibinfo  {journal} {JCAP}\ }\textbf {\bibinfo {volume} {1302}},\ \bibinfo
  {pages} {026}},\ \Eprint {https://arxiv.org/abs/1301.6637} {arXiv:1301.6637
  [astro-ph.HE]} \BibitemShut {NoStop}%
\bibitem [{\citenamefont {{A. Aab et al.}}(2014{\natexlab{a}})}]{Aab+14a}%
  \BibitemOpen
  \bibfield  {author} {\bibinfo {author} {\bibnamefont {{A. Aab et al.}}}
  (\bibinfo {collaboration} {Pierre Auger}),\ }\href
  {https://doi.org/10.1103/PhysRevD.90.122005} {\bibfield  {journal} {\bibinfo
  {journal} {Phys. Rev. D}\ }\textbf {\bibinfo {volume} {90}},\ \bibinfo
  {pages} {122005} (\bibinfo {year} {2014}{\natexlab{a}})},\ \Eprint
  {https://arxiv.org/abs/1409.4809} {arXiv:1409.4809 [astro-ph.HE]}
  \BibitemShut {NoStop}%
\bibitem [{\citenamefont {{A. Aab et al.}}(2014{\natexlab{b}})}]{Aab+14b}%
  \BibitemOpen
  \bibfield  {author} {\bibinfo {author} {\bibnamefont {{A. Aab et al.}}}
  (\bibinfo {collaboration} {Pierre Auger}),\ }\href
  {https://doi.org/10.1103/PhysRevD.90.122006} {\bibfield  {journal} {\bibinfo
  {journal} {Phys. Rev. D}\ }\textbf {\bibinfo {volume} {90}},\ \bibinfo
  {pages} {122006} (\bibinfo {year} {2014}{\natexlab{b}})},\ \Eprint
  {https://arxiv.org/abs/1409.5083} {arXiv:1409.5083 [astro-ph.HE]}
  \BibitemShut {NoStop}%
\bibitem [{\citenamefont {{V. Verzi, for the Pierre Auger
  Collaboration}}(2013)}]{Verzi13}%
  \BibitemOpen
  \bibfield  {author} {\bibinfo {author} {\bibnamefont {{V. Verzi, for the
  Pierre Auger Collaboration}}},\ }in\ \href@noop {} {\emph {\bibinfo
  {booktitle} {{33rd International Cosmic Ray Conference}}}}\ (\bibinfo {year}
  {2013})\ \Eprint {https://arxiv.org/abs/1307.5059} {arXiv:1307.5059
  [astro-ph.HE]} \BibitemShut {NoStop}%
\bibitem [{\citenamefont {{S. Baker and R. D. Cousins}}(1984)}]{Baker+83}%
  \BibitemOpen
  \bibfield  {author} {\bibinfo {author} {\bibnamefont {{S. Baker and R. D.
  Cousins}}},\ }\href {https://doi.org/10.1016/0167-5087(84)90016-4} {\bibfield
   {journal} {\bibinfo  {journal} {Nucl. Instrum. Meth.}\ }\textbf {\bibinfo
  {volume} {221}},\ \bibinfo {pages} {437} (\bibinfo {year}
  {1984})}\BibitemShut {NoStop}%
\bibitem [{\citenamefont {Zyla}\ \emph {et~al.}(2020)\citenamefont {Zyla} \emph
  {et~al.}}]{PDG}%
  \BibitemOpen
  \bibfield  {author} {\bibinfo {author} {\bibfnamefont {P.~A.}\ \bibnamefont
  {Zyla}} \emph {et~al.} (\bibinfo {collaboration} {Particle Data Group}),\
  }\href {https://doi.org/10.1093/ptep/ptaa104} {\bibfield  {journal} {\bibinfo
   {journal} {PTEP}\ }\textbf {\bibinfo {volume} {2020}},\ \bibinfo {pages}
  {083C01} (\bibinfo {year} {2020})}\BibitemShut {NoStop}%
\bibitem [{\citenamefont {{A. H. Rosenfeld}}(1975)}]{Rosenfeld75}%
  \BibitemOpen
  \bibfield  {author} {\bibinfo {author} {\bibnamefont {{A. H. Rosenfeld}}},\
  }\href {https://doi.org/10.1146/annurev.ns.25.120175.003011} {\bibfield
  {journal} {\bibinfo  {journal} {Ann. Rev. Nucl. Part. Sci.}\ }\textbf
  {\bibinfo {volume} {25}},\ \bibinfo {pages} {555} (\bibinfo {year}
  {1975})}\BibitemShut {NoStop}%
\bibitem [{\citenamefont {{G. J. Feldman and R. D.
  Cousins}}(1998)}]{Feldman+97}%
  \BibitemOpen
  \bibfield  {author} {\bibinfo {author} {\bibnamefont {{G. J. Feldman and R.
  D. Cousins}}},\ }\href {https://doi.org/10.1103/PhysRevD.57.3873} {\bibfield
  {journal} {\bibinfo  {journal} {Phys. Rev. D}\ }\textbf {\bibinfo {volume}
  {57}},\ \bibinfo {pages} {3873} (\bibinfo {year} {1998})},\ \Eprint
  {https://arxiv.org/abs/physics/9711021} {arXiv:physics/9711021
  [physics.data-an]} \BibitemShut {NoStop}%
\bibitem [{\citenamefont {{M. G. Aartsen et
  al.}}(2019)}]{IceCubeInelasticity19}%
  \BibitemOpen
  \bibfield  {author} {\bibinfo {author} {\bibnamefont {{M. G. Aartsen et
  al.}}} (\bibinfo {collaboration} {IceCube}),\ }\href
  {https://doi.org/10.1103/PhysRevD.99.032004} {\bibfield  {journal} {\bibinfo
  {journal} {Phys. Rev. D}\ }\textbf {\bibinfo {volume} {99}},\ \bibinfo
  {pages} {032004} (\bibinfo {year} {2019})},\ \Eprint
  {https://arxiv.org/abs/1808.07629} {arXiv:1808.07629 [hep-ex]} \BibitemShut
  {NoStop}%
\bibitem [{\citenamefont {{R. Abbasi et al.}}(2020)}]{IceCubeHESE20}%
  \BibitemOpen
  \bibfield  {author} {\bibinfo {author} {\bibnamefont {{R. Abbasi et al.}}}
  (\bibinfo {collaboration} {IceCube}),\ }\href@noop {} {\bibinfo {title} {{The
  IceCube high-energy starting event sample: Description and flux
  characterization with 7.5 years of data}}} (\bibinfo {year} {2020}),\ \Eprint
  {https://arxiv.org/abs/2011.03545} {arXiv:2011.03545 [astro-ph.HE]}
  \BibitemShut {NoStop}%
\end{thebibliography}%

\appendix 

\section{Single-component astrophysical neutrino flux}

\par
\begin{figure*}[htbp!]
	\centering
    \includegraphics[width=\textwidth]{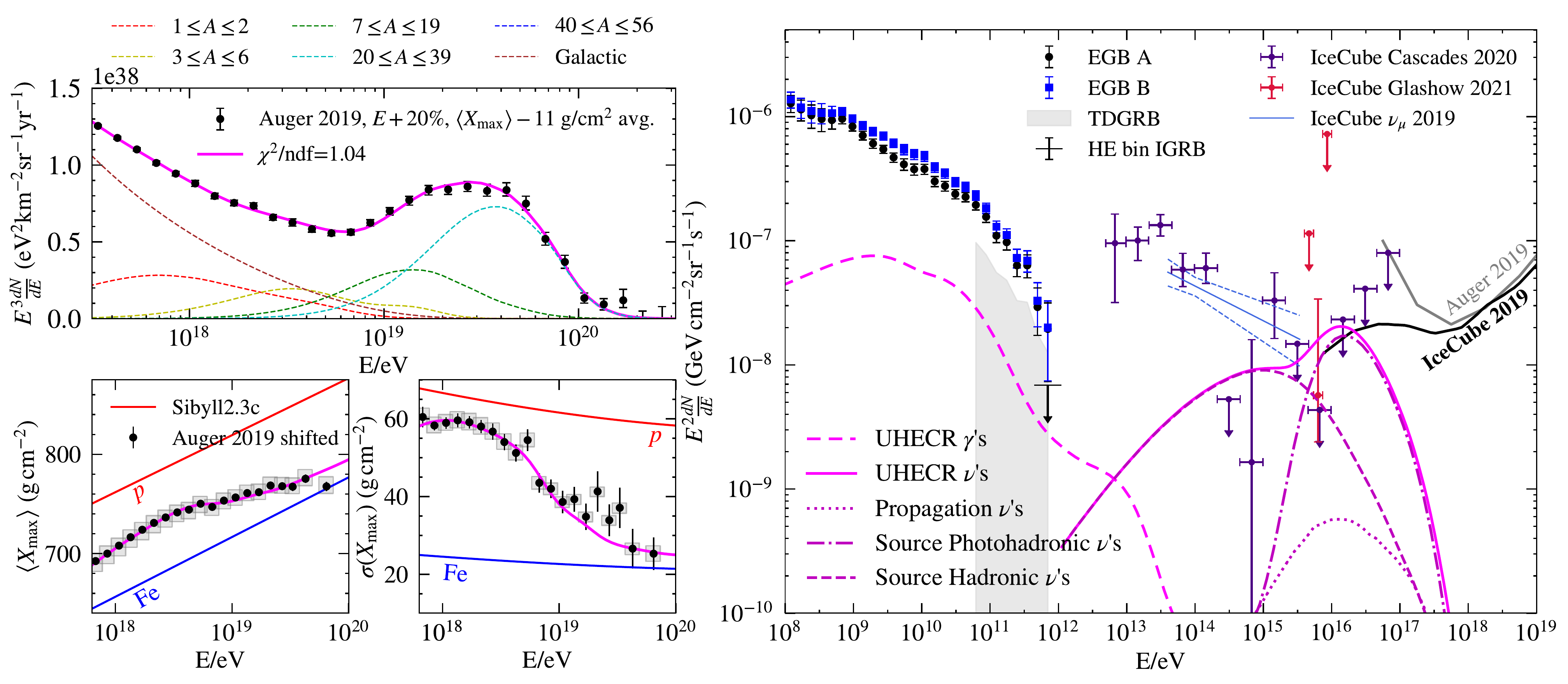}
	\caption{Predictions of the UHECR source model producing the best description of the astrophysical neutrino flux, when omitting a non-UHECR neutrino component; otherwise as in Fig. \ref{fig:astroNuBestFits_sibyll}.}
	\label{fig:astroNuBestFits_nolowENu_sibyll}
\end{figure*}
\begin{figure*}[htbp!]
	\centering
    \includegraphics[width=\textwidth]{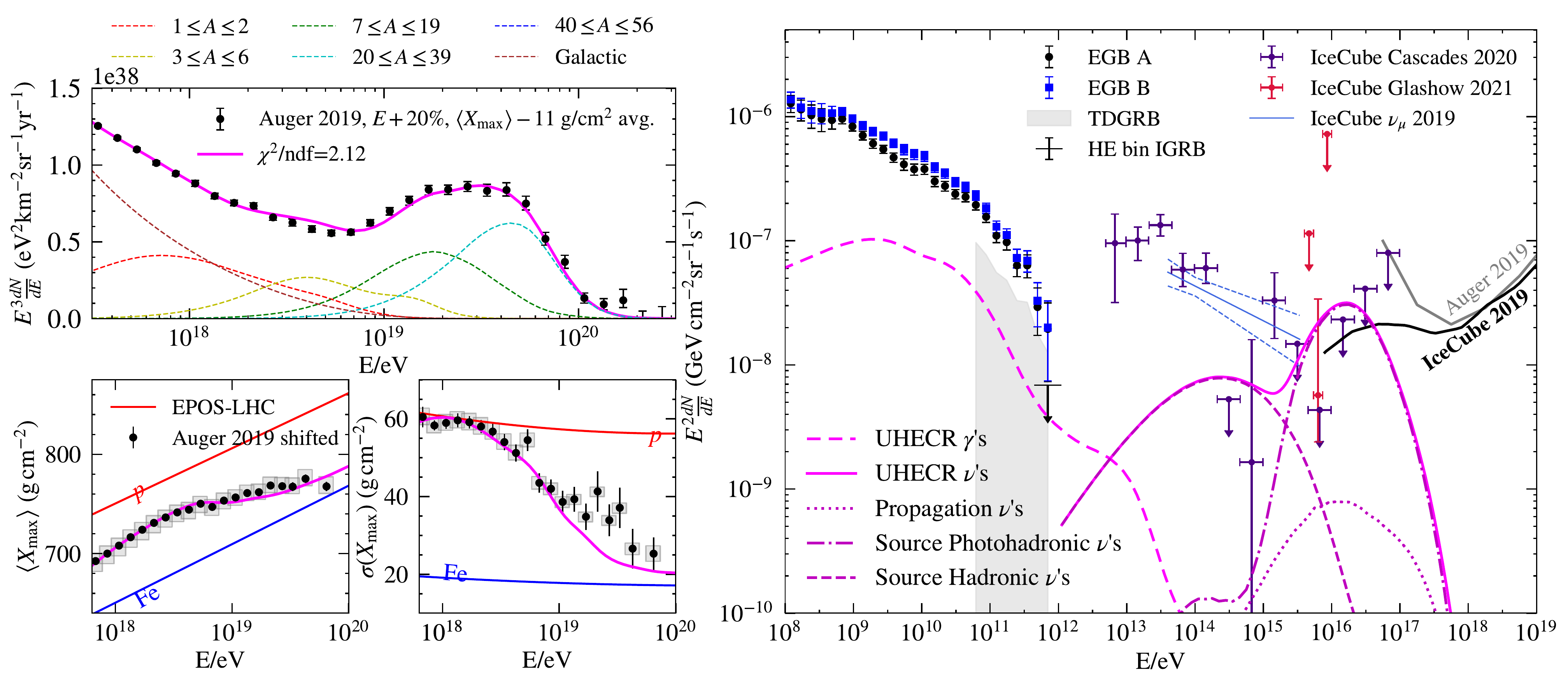}
	\caption{As in Fig.~\ref{fig:astroNuBestFits_nolowENu_sibyll} but for \textsc{EPOS-LHC}.} 
	\label{fig:astroNuBestFits_nolowENu_epos}
\end{figure*}

Figs. \ref{fig:astroNuBestFits_nolowENu_sibyll} and \ref{fig:astroNuBestFits_nolowENu_epos} show the best-fitting astrophysical neutrino flux produced by UHECR sources alone (i.e. not including a non-UHECR component of neutrinos as was explored in Section \ref{subsec:hiENus}). In particular, these fits cannot reproduce the low-energy part of the astrophysical neutrino flux observed by IceCube. This makes clear that UHECR sources alone cannot explain the entire flux of astrophysical neutrinos while also accurately reproducing the UHECR spectrum and composition and remaining consistent with bounds on EHE neutrinos.

\section{Softest spectral index $\gamma_\mathrm{inj} \lesssim -2$ compatible with multimessenger data}

\par
\begin{figure*}[htbp!]
	\centering
    \includegraphics[width=\textwidth]{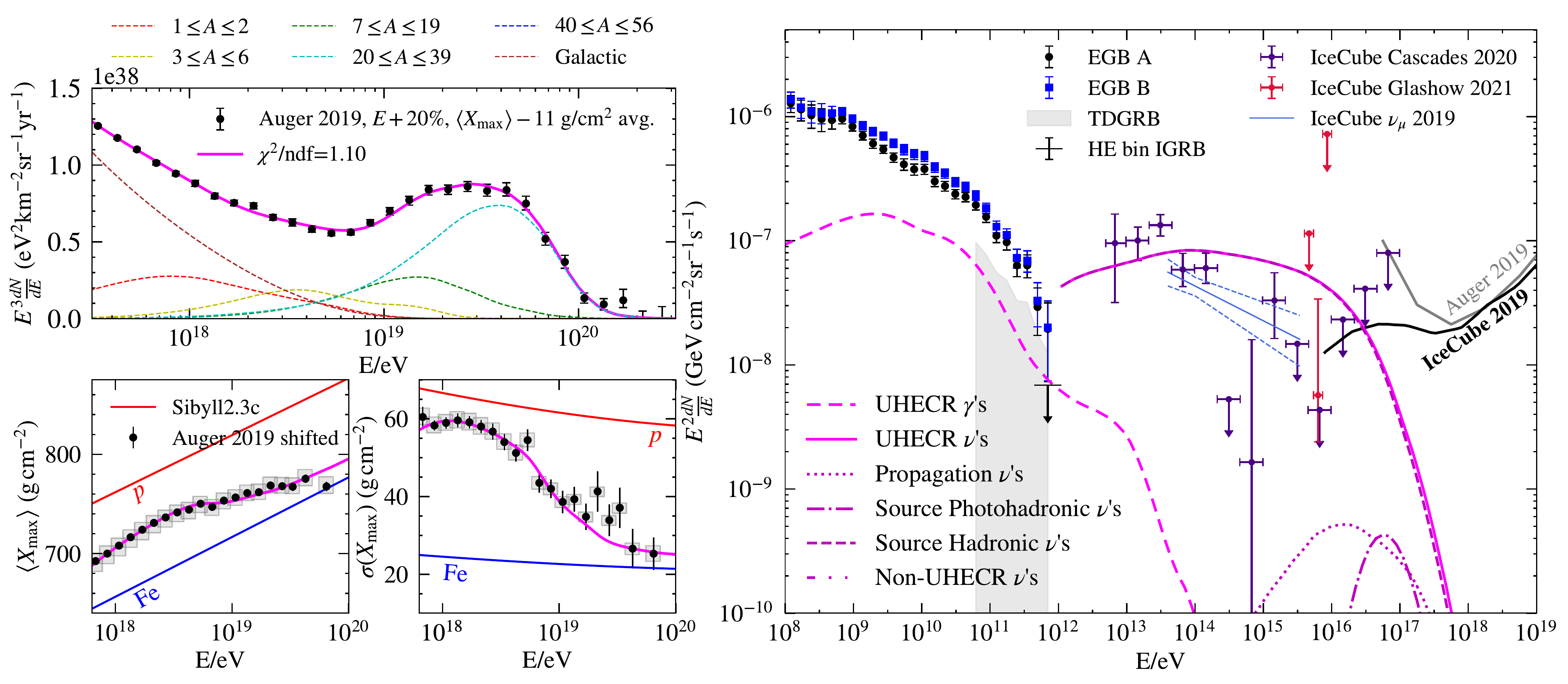}
	\caption{Predictions of the UHECR source model with the softest accelerator spectral index $\gamma_\mathrm{inj} \lesssim -2$ compatible with multimessenger data. This fit assumes \textsc{Sibyll2.3c}. A non-UHECR component of astrophysical neutrinos has been omitted. \textbf{Left:} Same as in Fig. \ref{fig:astroNuBestFits_sibyll}. \textbf{Right:} Same as in Fig. \ref{fig:astroNuBestFits_sibyll}.}
	\label{fig:softestIndex_sibyll}
\end{figure*}

Fig. \ref{fig:softestIndex_sibyll} shows the predictions of the UHECR model with the softest accelerator spectral index $\gamma_\mathrm{inj} \lesssim -2$ compatible with multimessenger data. This fit assumes the \textsc{Sibyll2.3c} HIM. (No corresponding fit is presented for \textsc{EPOS-LHC} as such soft spectral indices are entirely ruled out by UHECR and EHE neutrino constraints.) Table \ref{tab:softestIndex} gives the parameters for this particular fit which has spectral index $\gamma_\mathrm{inj} = -2.1$, similar to the spectral index predicted by diffusive shock acceleration. To achieve this fit a source size-to-magnetic coherence length ratio of $\sim 3000$ is required so that CRs at the highest rigidities enter the quasiballistic regime, allowing for a sufficient hardening of the spectrum.

\par
The right panel of Fig. \ref{fig:softestIndex_sibyll} shows that a soft spectral index $\gamma_\mathrm{inj} \sim -2$ can account for the general level of flux observed in astrophysical neutrinos at all energies, but not the shape. The observed astrophysical neutrino flux has an inferred spectral index of $\gamma \sim -2.5$ using a single power-law spectrum. However since the spectral index of neutrinos is inherited from the spectral index of the accelerated CRs, this is too soft to be compatible with multimessenger data.

\begin{table}[!htpb]

\centering
\setlength\tabcolsep{35pt}

\begin{tabularx}{\linewidth}{l S}

\hline \hline
\textbf{Parameter} & \textbf{\textsc{Sibyll2.3c}} \\
\hline
$\gamma_\mathrm{inj}$ & -2.1 \\
$\log_{10}(R_\mathrm{max}/\mathrm{V})$ & 18.8 \\
$\log_{10}{r_\mathrm{esc}}$ & 2.75 \\
$\log_{10}{r_{g\gamma}}$ & 0.45 \\
$\log_{10}(R_\mathrm{diff}/\mathrm{V})$ & 16.97 \\
$\log_{10}{r_\mathrm{size}}$ & 3.51 \\
$T/\mathrm{K}$ & 200 \\
$A_\mathrm{inj}$ & 30.1 \\
$f_\mathrm{gal}$ & 0.82 \\
$\gamma_\mathrm{gal}$ & -3.42 \\
$\log_{10}(E^\mathrm{galFe}_\mathrm{max}/\mathrm{eV})$ & 18.88 \\
$A_\mathrm{gal}$ & 26.3 \\
\hline
\end{tabularx}
\caption{\label{tab:softestIndex} Parameters corresponding to the model in Fig. \ref{fig:softestIndex_sibyll}, having the softest spectral index $\gamma_\mathrm{inj} \lesssim -2$ of the accelerator compatible with multimessenger data. See the caption of Table \ref{tab:astroNuBestFits} for parameter definitions.}

\end{table}

\end{document}